\documentclass{llncs}
\usepackage{amsmath}
\usepackage{graphicx}
\usepackage[T1]{fontenc}
\usepackage[latin1]{inputenc}

\begin{document}

\title{Automatic Summarization System coupled with a Question-Answering System (QAAS)}

\titlerunning{Automatic Summarization coupled with a Question Answering}

\author{Juan-Manuel Torres-Moreno\inst{1,2}, Pier-Luc St-Onge, Michel Gagnon\inst{2}, Marc El-Bèze\inst{1} and Patrice Bellot\inst{1} }

\authorrunning{Torres-Moreno et al.} 

\institute{Laboratoire Informatique d'Avignon \\
339 chemin des Meinajari\`es, BP1228, 84911 Avignon Cedex 9, France. \\
\email{juan-manuel.torres@univ-avignon.fr }  \vspace*{0.5cm}
\and \'Ecole Polytechnique de Montréal\\
CP 6079 Succ. Centre-ville, H3C 3A7 Montr\'eal (Qu\'ebec), Canada.
}


\maketitle

\begin{abstract}
  \textsc{Cortex} is an automatic generic document summarization system.
  To select the most relevant sentences of a document, it uses an optimal decision algorithm that combines several metrics. 
  The metrics processes, weighting and extract pertinence sentences by statistical and informational algorithms. 
  This technique might improve a Question-Answering system, whose function is to provide an exact answer to a question in natural language. 
  In this paper, we   present the results obtained by coupling the {\sc Cortex} summarizer with
  a Question-Answering system (QAAS). Two configurations have been evaluated. In the
  first one, a low compression level is selected and the summarization
  system is only used as a noise filter. In the second configuration, the system
  actually functions as a summarizer, with a very high level of
  compression. 
  Our results on French corpus demonstrate that the coupling of Automatic Summarization system with a Question-Answering system is promising.
  Then the system has been adapted to generate a customized
  summary depending on the specific question. 
  Tests on a french multi-document corpus have been realized, and the personalized QAAS system obtains the best performances.
\end{abstract}


\section{Introduction}

Automatic summarization is indispensable to cope with ever increasing volumes of valuable information. An abstract is by far the most concrete and most recognized kind of text condensation \cite{ANS79}.
We adopted a simpler method, usually called \textit{extraction}, that allow to generate summaries by extraction of pertinence sentences \cite{luhn1958,edmundson1969}. 
Essentially, extracting aims at producing a shorter version of the text by selecting the most relevant sentences of the original text, which we juxtapose without any modification.
Linguistic methods, notably semantic analysis, are relevant, but
their application remains difficult or limited to restricted
domains \cite{SAG00,Teufel&Moens02}. 
The vector space model \cite{SAL71,SAL83} has been used in information extraction, information retrieval, question-answering,  and it may also be used in text summarization. 
Furthermore, statistical, neural, SVM and connexionist methods are often employed in several areas of text processing \cite{THO02,DEE90,VER91,BAL96,MEU97}. 
Actually, the existing techniques only allow to produce summaries of the informative type \cite{MOR99}. 
Our research tries to generate this kind of summaries.
{\sc Cortex}\footnote{\textsl{COndensés et Résumés de TEXte} (Text Condensation and Summarization).} is an
automatic summarization system, recently developed \cite{TOR04} which combines
several statistical methods with an optimal decision algorithm, to
choose the most relevant sentences.

An open domain Question-Answering system (QA) has to precisely answer a question expressed in natural language. 
QA systems are confronted with a fine and difficult task because they are expected to supply specific information and not whole documents. 
At present there exists a strong demand for this kind of text processing systems on the Internet. 
A QA system comprises, \emph{a priori}, the following stages \cite{JAC00}:
\begin{itemize}
 \item Transform the questions into queries, then associate them to a set of documents;
 \item Filter and sort these documents to calculate various degrees of similarity;
 \item Identify the sentences which might contain the answers, then extract text fragments from them which constitute the answers. In this phase an analysis using Named Entities (NE) is essential to find the expected answers.
\end{itemize}

Most research efforts in summarization emphasize generic summarization \cite{Abracos&Lopes97,Teufel&Moens97,Hovy&Lin99}.
User query terms are commonly used in information retrieval tasks.
However, there are few  papers in literature that propose to employ this approach 
in summarization systems \cite{Kupiec&al.95,Tombros&al.98,SCH01}.
In the systems described in \cite{Kupiec&al.95}, a learning approach is used (performed).
A document set is used to train a classifier that estimates the probability that a given sentence 
is included in the extract. In \cite{Tombros&al.98}, several features (document title, location of a sentence in the document,
cluster of significant words and occurrence of terms present in the query) are applied to score the sentences.
In \cite{SCH01} learning and feature approches are combined in a two step system: a training system and a generator system. 
Score features include short length sentence, sentence position in the document, sentence position in the paragraph, and tf.idf
metrics. Our generic summarization system includes a set of ten independent metrics combined by a Decision Algorithm. Query-based
summaries can be generated by our system using a modification of the scoring method. In both cases, no training phase is necessary in our system.

In this paper we present the coupling of an algorithm of automatic summarization with a Question-Answering system, which allows to decrease the document search  space and to increase the number of correct answers returned by the system. 
Two scenarios have been evaluated: 
in the first one the summarization process is used as a noise filter (it condenses texts at a low compression rate), and in the second one as a true summarization system (it condenses at high rates). 
In Section 2, the preprocessing technique is presented. 
In Section 3, the {\sc Cortex} algorithm is described: several metrics and a Decision Algorithm (DA) are
presented. In Section 4, we analyze the sensibility of metrics and DA. 
In Section 5, two main evaluation methods are described and applied. 
In Section 6 end 7, experiments and results of applying both {\sc Cortex} and QA systems are described. Finally, some conclusions and future work are presented.

\section{Pre-processing}

We process texts according to the vector space model \cite{CHR99}, a text representation very different from linguistic structural
analysis, but which allows to efficiently process large volumes of documents \cite{SAL71,MOR99}. 
Texts are represented in a vector space to which several classic numeric algorithms are applied. 

\begin{description}
\item[Filtering]

In a first step, the {\sc Cortex} algorithm pre-processes each text in the corpus. 
The original text contains $N_{W}$ words which can be function words (articles, prepositions, adjectives, adverbs), nouns and conjugated verbs, but also compound words which often represent a very
specific concept. 
All these words may occur repeatedly. 
It is important to decide whether to utilize inflected forms or base forms.
That is why we prefer the more abstract notion of {\it term} instead of {\it word} \cite{CHR99}. 
To reduce the complexity of the text, various filters are applied to the lexicon: 
the (optional) deletion of function words and auxiliary verbs\footnote{Our exemples and tests in this paper are all in French: \emph{être} (to be),
  \emph{avoir} (to have), \emph{pouvoir} (can), \emph{devoir}
  (must)...}, common expressions\footnote{\emph{par exemple} (for
  example), \emph{ceci} (that is), \emph{chacun/chacune} (each of)
  ...}, text in parentheses (which often contains additional
information which is not essential for the general comprehension),
numbers (numeric and/or textual)\footnote{In the case of generic abstracts, we decided to delete
numbers. However, when coupled with the QA system, we do not delete them, because
the answers are often of numerical type.} and symbols, such as
$\left\langle\$\right\rangle$, $\left\langle\#\right\rangle$,
$\left\langle*\right\rangle$, etc. In this stage, we employ several
negative dictionaries or generic stoplists.

\item[Lemmatization and Stemming]

In morphologically rich languages, such as Romanic languages, it is
essential to lemmatize the words. This considerably reduces
the size of the lexicon. Simple lemmatization consists of finding
the lemma of the conjugated verbs and replacing the plural and/or
feminine words with the singular masculine form before counting the
number of occurrences. In this task, a dictionary containing
approximately 330,000 entries was used. After lemmatization, we
applied a stemming \cite{PAI90,POR80} affix removal algorithm (based on Porter's rules
\cite{POR80}) to obtain the stem of each lemma. Stemming (or
conflating) words allows to reduce the morphological variants of the
words to their stem \cite{FIA00}. In these processes, it is assumed
that words semantically related have the same stem.  So the words
{\it chante, chantaient,
  chanté, chanteront} and eventually {\it chanteur}\footnote{Respectively {\bfseries sing}, {\bfseries sang}, {\bfseries sung}, (will) {\bfseries sing} and {\bfseries singer}.} are transformed to
the same form {\bf chanter} (to sing).  This twofold process decreases
the curse of dimensionality, which causes severe problems of matrix
representation for large data volumes. Lemmatization/stemming
identifies a number of terms which defines the dimensions of the
vector space. Some additional mechanisms to decrease the size of the
lexicon are also applied.  One of them is compound words detection.
Compound words are found, then transformed into a unique
lemmatized/stemmed term\footnote{\emph{pomme de terre}/\emph{pommes
    de terre} becomes \textbf{pomme\_de\_terre} (potato).}. We also
investigate other methods for lexicon reduction, for example by
grouping synonyms by means of specialized dictionaries.

\item[Split sentences]

Given the cognitive nature of summaries, we split the texts into variable length segments (sentences), according to one or more suitable 
criteria\footnote{In the case of summarization of very long texts, it is more suitable to apply a segmentation by paragraph or page 
rather than by sentences.}. 
Fixed size segmentation was ruled out, because we want to extract complete sentences. 
Period $\left\langle.\right\rangle$, carriage return $\left\langle \hookleftarrow CR\right\rangle$,
colon $\left\langle :\right\rangle$, question mark $\left\langle ?\right\rangle$ and
exclamation mark $\left\langle !\right\rangle$ (or their combinations) may be taken as sentence
delimiters. 
Since electronic addresses and Internet sites (URLs) always contain periods, it is essential to detect and transform them in this phase.


\item[Title detection]

The titles (document title and section titles) found in a document are very informative. However, in raw texts, the title is not marked explicitly. 
Therefore, to detect it, some heuristics are needed.
Conceptually, the title can be processed as a particular segment. 
A segment is declared to be the "main-title" following the rules below:
\begin{itemize}
  \item The words of the first sentence are in capital letters.
  \item The first sentence.
  \item The first sentence is separated from the text by a carriage return.
  \item The 10 first words of a text.
\end{itemize}
At the end of these processes, an XML file with a simple structure is obtained: 

{\scriptsize
\begin{verbatim}
 <?xml version="1.0" encoding="UTF-8" ?>
 <Texte Langue="Fra" Title="This is the title">
   <S> Text for processing </S>
      <Subtitle_1> Title of section 1 </Subtitle_1>
         <S> Text to processing </S>
         <S> Text to processing </S>
         ...
      <Subtitle_2> Title of section 2 </Subtitle_2>
         ...
     <S> More text to processing </S>
     ...
 </Text>
\end{verbatim}
}


\begin{figure}
\centerline{\includegraphics[width=25pc]{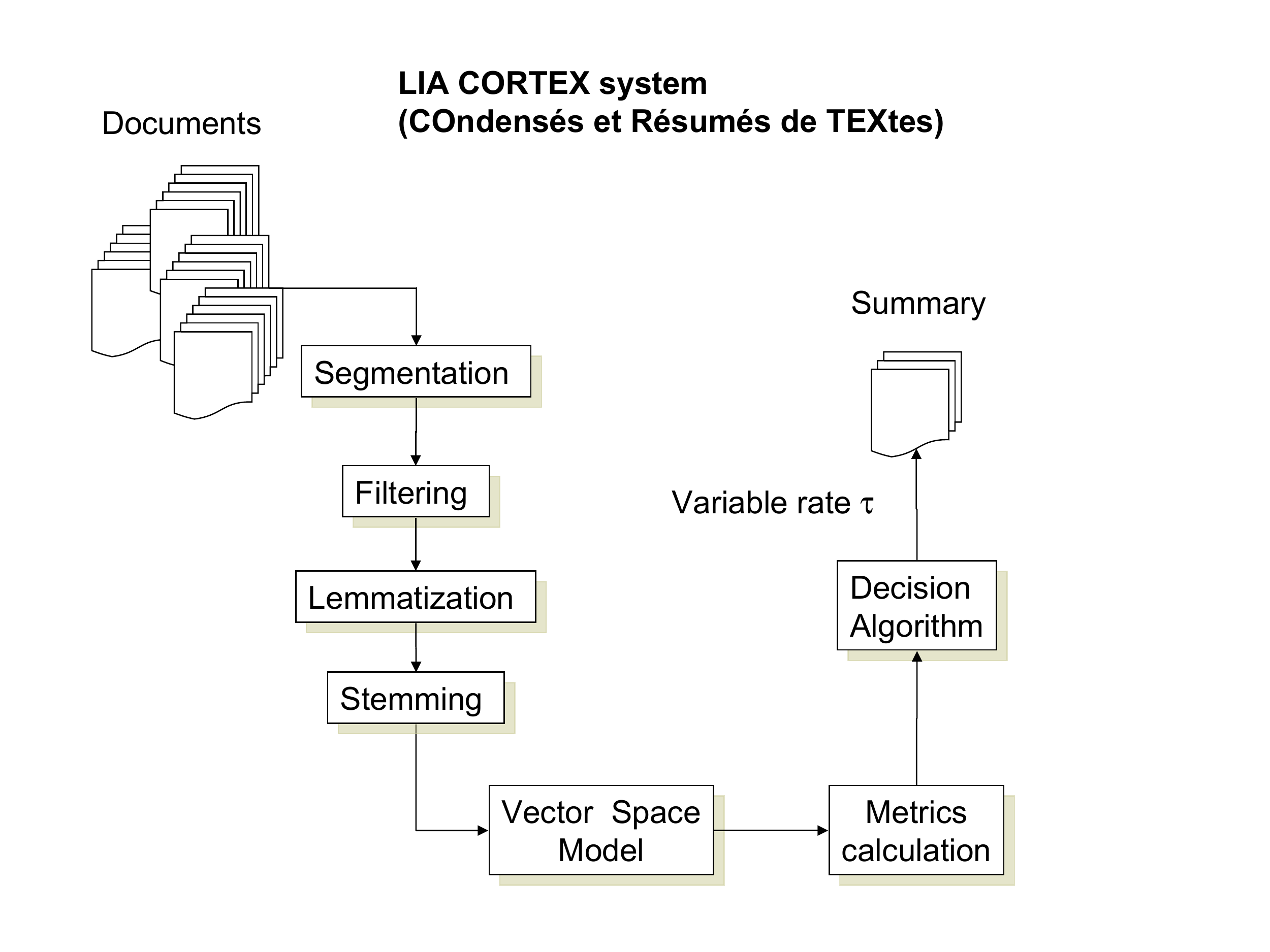}}
 \caption{\em \small General architecture of Automatic Summarization system LIA-{\sc Cortex}.}
  \label{fig:lia-qaas}
\end{figure}

\noindent 
After pre-processing, a text representation in Vector Space Model is constructed. 
Then, we apply several statistical processes to score sentences.
The summary is generated by selecting the sentences with higher scores.  
In Figure \ref{fig:lia-qaas}, we present a diagram of the {\sc Cortex} system, developed at Laboratoire Informatique d'Avignon\footnote{\url{http://www.lia.univ-avignon.fr}} (LIA). 
In the next section, we present the weigthing sentences algorithm of {\sc Cortex}.

\end{description}

\section{The {\sc Cortex} algorithm}\label{Presentation}

In this section, the matrices, the metrics and the Decision Algorithm of the {\sc Cortex} system will be described. 
After the pre-processor has filtered the text and lemmatized the words (to group those of 
the same family) the selection of relevant sentences can be
started. For every sentence, the metrics, which are all based on
the matrices of either presence or frequency of terms, are
calculated and combined by the Decision Algorithm described later
(see Section \ref{AlgoDecision}). The sentences are then ranked
according to the values obtained. Depending on the desired
compression rate, the sorted sentences will be used to produce the summary.
We define the following variables:
        \begin{description}
            \item[$N_{W}$ :]
                Total number of different raw words.
            \item[$N_{S}$ :]
                Total number of sentences.
            \item[$N_{T}$ :]
                Total number of titles (document title and section titles) in the text.
            \item[$N_{M}$ :]
                Number of different terms remaining after filtering.
            \item[$N_{L}$ :]
                Size of the "relevant" lexicon, i.e. the number of words appearing at least twice in the text.
        \end{description}

Based on the terms that remain in the text after filtering, a
frequency matrix ${\gamma}$ is constructed in the following way: every element
$\gamma_i^{\mu}$ of this matrix represents the number of
occurrences of the word $i$ in the sentence $\mu$.

\begin{equation}\label{MatriceFreq}
   \mathbf{\gamma} =
    \left[
    \begin{array}{ccccccc}
      \gamma^{1}_{1} &        \gamma^{1}_{2} &             \ldots &    \gamma^{1}_{i} &        \ldots &    \gamma^{1}_{N_{L}}      \\
      \gamma^{2}_{1} &        \gamma^{2}_{2} &             \ldots &    \gamma^{2}_{i} &        \ldots &    \gamma^{2}_{N_{L}}      \\
      \vdots &                \vdots &                     \ddots &    \vdots &                \ddots &    \vdots                  \\
      \gamma^{\mu}_{1} &      \gamma^{\mu}_{2} &           \ldots &    \gamma^{\mu}_{i} &      \ldots &    \gamma^{\mu}_{N_{L}}    \\
      \vdots &                \vdots &                     \ddots &    \vdots &                \ddots &    \vdots                  \\
      \gamma^{N_{S}}_{1} &    \gamma^{N_{S}}_{2} &        \ldots &    \gamma^{N_{S}}_{i} &    \ldots &    \gamma^{N_{S}}_{N_{L}}
    \end{array}
    \right], \quad \gamma^{\mu}_{i} \in \{ 0, 1, 2, \dots \}
\end{equation}

\noindent Another matrix $\xi$, called {\sl binary virtual} or {\sl presence matrix}, is defined as:
\begin{equation}\label{MatriceBin}
      \xi^{\mu}_{i} = \left\{ \begin{array}{l} 1 \quad \textrm{if } \gamma^{\mu}_{i} \neq 0 \\ 0 \quad \textrm{elsewhere} \end{array}
      \right\}
\end{equation}

Every line of these matrices represents a sentence of the text.
The sentences are indexed by a value $\mu$ varying from 1 to
$N_S$. Every column represents a term of the text. Terms are indexed by a value $i$ varying from 1 to $N_L$. The titles are
stored in another matrix $\gamma^{T}$. Matrices $\bf\gamma$ and $\gamma^{T}$ are the
frequency matrix of the sentences, and frequency matrix of the titles, respectively.


Pre-processing phase transforms the text into a set of $N_S$ sentences or segments and  $N_L$ retained terms  which are regarded as relevant. 
The relation $N_L \le N_M \le N_W$ is always true. 
We define $\rho_L$ as:
\begin{equation}
   \rho_L = \frac{N_L}{N_W}
   \label{eq:reduccion}
\end{equation}


It is important to note that the matrices $\gamma$ and $\xi$ are very sparse
because  every line (representing a sentence) contains only a
small part of the vocabulary. Because of this, fast matrix
manipulation algorithms had to be adapted and implemented.
We estimated $\rho_L$ over 20,000 text documents (from the corpus \textsl{Le Monde} that
will be explained in section \ref{Sec:exp2} Experiments II). We obtained $\rho_L \approx 0.52$, on average.

To obtain the final summary, the user sets the compression ratio $\tau$ as a fraction (in percent) of the number $N_S$ of sentences, or the number $N_W$ of words.

\subsection{The metrics}

Important mathematical and statistical information can be gained
from the "term-segment" matrices  $\xi$ and $\gamma$, to be used
in the condensation process. In our experiments, $\Gamma=10$ metrics
were calculated (frequencies, entropy, Hamming and hybrid) based
on these matrices. The more relevant a segment is, the higher are
the values of its metrics. Subsequently, the $\Gamma$ metrics used are
explained:

\begin{enumerate}
    \item \textbf{ Frequency measures.}
    \begin{enumerate}
                \item Term Frequency $F$:
                    The Term Frequency metrics \cite{SAL89} counts the number of relevant words in every sentence $\mu$.
                    Thus, if a sentence contains more important words, it has more chances to be retained.
                    If the sentence is longer, it usually includes more relevant words,
                    thus it has a bigger chance to be retained.
                    Consequently, the summaries generated based on this metrics (generally) contain the long sentences
                    \footnote{It is important to note that in our context, metrics $F$ (and below entropy $E$) is useful 
                    only after the filtering/lemmatisation processes: the function words and words with $F<1$ are not present
                    in the lexicon of $N_L$ words.}.
                    \begin{equation}
                        F^{\mu} = \sum^{N_L}_{i = 1} \gamma^{\mu}_{i}
                    \end{equation}
                    Note that we can easily calculate $T$, the total number of terms occurring in the text after filtering:
                    \begin{equation}
                       T = \sum^{N_{S}}_{\mu = 1} \sum^{N_{L}}_{i = 1} \gamma^{\mu}_{i} = \sum^{N_{S}}_{\mu = 1} F^{\mu}
                    \end{equation}
               \item Interactivity of segments $I$: The {\sc Cortex}
                    system exploits the existence of a network of
                    words of the same family present in several
                    sentences.  For every distinct term in a sentence, we count the number
                    of sentences, except the current sentence,
                    containing this word\footnote{For example, if the word \textsl{"aimer"} ("to love")
                    occurs twice in a sentence, that accounts for a single
                           "distinct" word. For this reason, we use the
                           matrix of presences $\xi$.}. 
                    Then the current sentence $\mu$ is said to be in interactivity with $N_i$ sentences
                    by the word $i$.  The $N_i$ value of all  words
                    in the sentence are added to obtain their weights.
                    \begin{equation}
                        I^{\mu} = \sum^{N_L}_{\begin{subarray}{c} i = 1 \\ \xi^{\mu}_{i} \neq 0 \end{subarray}}
                                  \sum^{N_S}_{\begin{subarray}{c} j = 1 \\ j \neq \mu \end{subarray}} \xi^{j}_{i}
                    \end{equation}
                \item Sum of probability frequencies $\Delta$:
                     This metrics balances the frequency of the words in
                     the sentences according to their global frequency:
                     \begin{equation}
                        \Delta^{\mu} = \sum^{N_L}_{i = 1} p_{i} \gamma^{\mu}_{i}
                     \end{equation}
                     With:
                     \begin{equation}
                        p_{i} = \frac{1}{T} \sum^{N_S}_{\mu = 1} \gamma^{\mu}_{i}
                     \end{equation}
                     The values $p_i$ are the probabilities of occurrence of term $i$ in the text.
                     The more often a word (or a family of words) occurs in a text, the greater will be its weight
                     in the sentences.
                     The product $p_i \gamma^{\mu}_{i}$ of metrics $\Delta$ is not similar to tf.idf 
                     (Term frequency - Inverse document frequency \cite{SAL89}) weigthing: 
                     the $p_i$ are values de probability of a term $i$ in all segments, 
                     instead of inverse document frequencies, and no logarithm or
                     square function is used in calculations.
    \end{enumerate}
    \item \textbf{ Entropy.} The entropy $E$ is another measure depending on the probability of a word
                            in a text. If the probability $p_{i}$ of a word is high, then the sentences
                             which contain this word may be favoured:
                    \begin{equation}
                        E^{\mu} = -\sum^{N_L}_{\begin{subarray}{c} i=1 \\
                        \xi^{\mu}_{i} \neq 0 \end{subarray}} p_{i} \log_{2} p_{i}
                    \end{equation}

     \item \textbf{ Measures of Hamming.}
                    These metrics use a Hamming matrix $H$, a square matrix ${N_L} \times {N_L}$, defined in the following way: 
                    \begin{equation}
                       H^{m}_{n}=\sum^{N_S}_{j=1} \left\{\begin{array}{l} 1 \quad \textrm{if } \xi^{j}_{m} \neq \xi^{j}_{n} \\ 0
                                    \quad \textrm{elsewhere} \end{array} \right\}
                                    \quad \textrm{for } \begin{array}{l} m \in \left[ 2, N_L \right] \\
                                    n \in \left[ 1, m \right] \end{array}
                    \end{equation}
                    The Hamming matrix is a lower triangular matrix where the index $m$ represents the
                    line and the index $n$ the column, corresponding to the index of words, where $m > n$.

                    The idea is to identify the terms which are
                    semantically connected.  In this way, two terms
                    which might be synonyms will have a high value
                    in $H$ because we do not expect to find them in
                    the same sentence, i.e. this matrix represents 
                    the number of sentences that contains only one of two words but not both.\\

    \begin{enumerate}
        \item Hamming distances $\Psi$:
                    The main idea is that if two important words (maybe synonyms) are in the same sentence,
                    this sentence must certainly be important.
                    The importance of every pair of words directly corresponds to the value in the
                    Hamming matrix $H$\footnote{The sum of the Hamming distances is the most 
                    resource-intensive metrics to be calculated.
                    It takes more time than all other metrics combined because its complexity is $O(N_{L}^2)$.}.
                    The metrics of the Hamming distances is calculated as follows:
                    \begin{equation}
                        \Psi^{\mu} = \sum^{N_L}_{\begin{subarray}{c} m = 2 \\ \xi^{\mu}_{m} \neq 0 \end{subarray}}
                                    \sum^{m}_{\begin{subarray}{c} n=1 \\ \xi^{\mu}_{n} \neq 0 \end{subarray}} H^{m}_{n}
                    \end{equation}
                \item Hamming weight of segments $\phi$:
                    The Hamming weight of segments is similar to the metrics of frequencies $F$.
                    In fact, instead of adding the frequencies of a sentence, the occurrences $\xi$ are added.
                    Thus, a sentence with a large vocabulary  is favoured.
                    \begin{equation}
                        \phi^{\mu} = \sum^{N_L}_{i = 1} \xi^{\mu}_{i}
                    \end{equation}
                \item Sum of Hamming weight of words per segment $\Theta$: \label{PresenterT}
                    This metrics closely resembles the metrics of interactivity $I$.
                    The difference is that for every word present in a sentence $\mu$, all the
                    occurrences of this word in the text are counted and not only their presence
                    in all other sentences except the current sentence.
                    We thus obtain:
                    \begin{equation}
                        \Theta^{\mu} = \sum^{N_L}_{\begin{subarray}{c} i=1 \\
                        \xi^{\mu}_{i} \neq 0 \end{subarray}} \psi_{i}
                    \end{equation}
                    and $\psi_{i}$ as the sum of the occurrences of every word.
                    \begin{equation}
                        \psi_{i} = \sum^{N_S}_{\mu = 1} \xi^{\mu}_{i}
                    \end{equation}
                \item Hamming weight heavy $\Pi$: 
                    Among the sentences containing the same set of important words, 
                    how do we know which one is the best, i.e. which one of these sentences is the more informative?
                    The solution is to choose the one that contains the biggest part of the lexicon.
                    Already, the metrics $\Theta^{\mu}$ is relatively sensitive to the different words in a sentence.
                    However, if  this metrics is again multiplied by the number of different words in a sentence
                   ($\phi^{\mu}$), we are capable to identify the most informative sentences.
                    \begin{equation}
                        \Pi^{\mu} = \phi^{\mu} \Theta^{\mu}
                    \end{equation}
                \item Sum of Hamming weights of words by frequency $\Omega$: 
                    The sum of the Hamming weights of the words by frequencies uses the frequencies as
                    factor instead of the presence as in the case of the metrics $\Theta^{\mu}$.
                    The sentences containing the most important words several times will be favoured.
                    \begin{equation}
                        \Omega^{\mu} = \sum^{N_L}_{i = 1} \psi_{i} \gamma^{\mu}_{i}
                    \end{equation}
                    Note that $\psi_{i}$ has been calculated in the metrics $\Theta$, and
                    $\gamma^{\mu}_{i}$ represents the number of times that the term $i$ is present in the sentence $\mu$.
    \end{enumerate}

    \item \textbf{ Titles and subtitles.} Almost all the texts have a main title.
                    Some also have subtitles. So, important information can be deduced from
                    the document structure.
                    Angle between a title and a sentence $\theta$: 
                    The purpose of this metrics is to favor the sentences which refer to the subject in the title.
                    In fact, we compare, word by word, every sentence to the title\footnote{The metrics $\Theta$ will be
                    used to get personalized abstracts (see subsection \ref{Sect:personn}).} (main title or subtitle).
                    To combine the comparisons, we calculate the normalized  $N_L$ dimensional scalar vector
                    product between the sentence and the title vector $\gamma^{T}$, and finally the cosine of this value:
                    \begin{equation}
                    \label{eq:theta}
                        \theta^{\mu} = \cos \left( \frac{\sum^{N_L}_{i = 1}\gamma^{\mu}_{i}\gamma^{T_{\mu}}_{i}}{\left\Arrowvert
                        \gamma^{\mu} \right\Arrowvert \left\Arrowvert \gamma^{T_{\mu}} \right\Arrowvert} \right)
                    \end{equation}
 \end{enumerate}

\subsection{Normalization of the metrics}
Before using the $\Gamma$ metrics in the decision algorithm, they
have to be normalized. Therefore, every metrics is calculated for
all the sentences. The value $\lambda_{\mu}$ for a sentence $\mu=1, \cdots, N_S$ is shown below:
            \begin{equation}
                \left\Arrowvert \lambda_{\mu} \right\Arrowvert = \frac{\lambda_{\mu} - m}{M - m}
            \end{equation}
            where:
            \begin{displaymath}
                \begin{array}{rcl}
                    m & = & \min \left\{ \lambda_{j} \textrm{ for } j \in \left[ 1, N_S \right] \right\} \\
                    M & = & \max \left\{ \lambda_{j} \textrm{ for } j \in \left[ 1, N_S \right] \right\} \\
                \end{array}
            \end{displaymath}
            Every metrics normalized takes values in the range [0,1].

\subsection{Decision algorithm}
\label{AlgoDecision}

The Decision Algorithm (DA) combines all normalized metrics in a sophisticated way. 
Two averages are calculated: the positive tendency, that is
$\lambda_{\mu} > 0.5$, and the negative tendency, for
$\lambda_{\mu}< 0.5$ (the case $\lambda_{\mu} = 0.5$ is ignored).
To calculate this average, we always divide by the total number of
metrics $\Gamma$ and not by the number of "positive" or "negative"
elements (real average of the tendencies).

So, by dividing by $\Gamma$, we have developed an algorithm more decisive than the simple average
\footnote{Contrary to simple average, which may be ambiguous if the value is close to 0.5, our algorithm chooses to penalize the sentences with a score of exactly 0.5.} and even more realistic than the real average of the tendencies. Here is the decision algorithm that allows to include the vote of each metrics:
            \begin{eqnarray}
                \sum^{\mu} {\alpha} & = & \sum^{\Gamma}_{\begin{subarray}{c} \nu = 1
                \\ \left\Arrowvert \lambda^{\nu}_{\mu}
                \right\Arrowvert > 0.5 \end{subarray}} \left(\left\Arrowvert \lambda^{\nu}_{\mu} \right\Arrowvert - 0.5\right)
                \\ \sum^{\mu} {\beta}  & = & \sum^{\Gamma}_{\begin{subarray}{c} \nu = 1 \\ \left\Arrowvert \lambda^{\nu}_{\mu}
                \right\Arrowvert < 0.5 \end{subarray}} \left(0.5 - \left\Arrowvert \lambda^{\nu}_{\mu} \right\Arrowvert\right)
            \end{eqnarray}
$\nu$ is the index of the metrics, $\sum^{\Gamma}_{\nu}$ is the sum of the absolute differences between $\left\Arrowvert \lambda \right\Arrowvert$ and 0.5, $\sum^{\mu} \alpha$ are the "positive" normalized metrics,
$\sum^{\mu} \beta$ the negative normalized metrics and $\Gamma$ the number of metrics used.
The value attributed to every sentence is calculated in the following way:

 \begin{eqnarray*}
  & \mbox{If}& \left( \sum^{\mu}\alpha>\sum^{\mu}\beta \right) \\
  &          & \mbox{then } \Lambda^{\mu} = 0.5 + \frac{\sum^{\mu} {\alpha}}{\Gamma} \mbox{ : DA is chosen in order to advantage the segment $\mu$} \\
  &          & \mbox{else } \Lambda^{\mu} = 0.5 - \frac{\sum^{\mu} {\beta}}{\Gamma}  \mbox{ : DA is chosen in order to disadvantage it}
 \end{eqnarray*}

\noindent $\Lambda^{\mu}$ is the value to finally decide
whether or not to retain  the sentence $\mu$. In the end, $N_{S}$
sentences are sorted according to this value $\Lambda^{\mu};
\mu=1, \cdots, N_S$. The compression rate $\tau$ determines the
final number of sentences, which are chosen from the sorted list.

\section{Metrics sensibility}

\begin{figure}
 \centerline{\includegraphics[width=16pc]{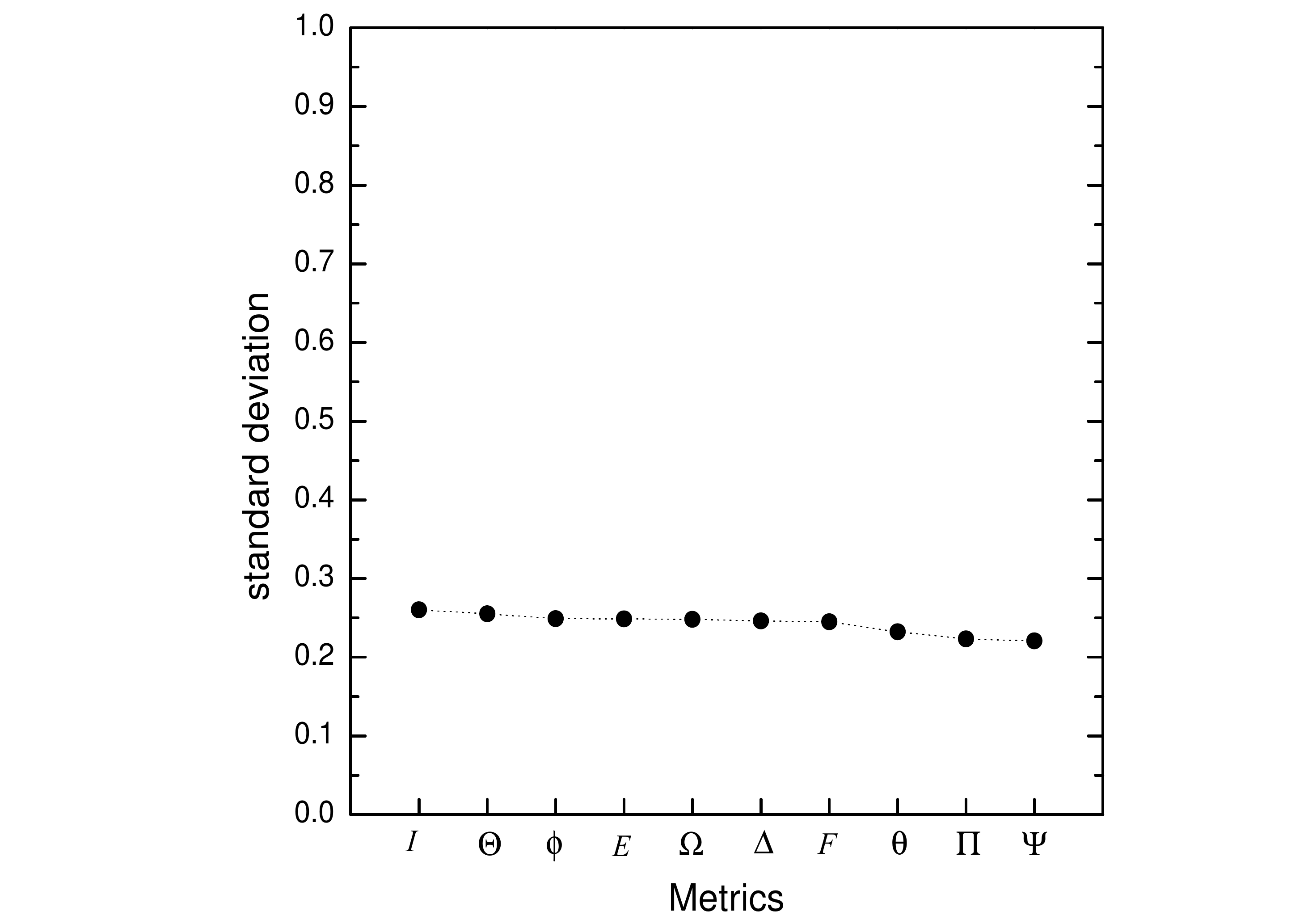}
             \includegraphics[width=16pc]{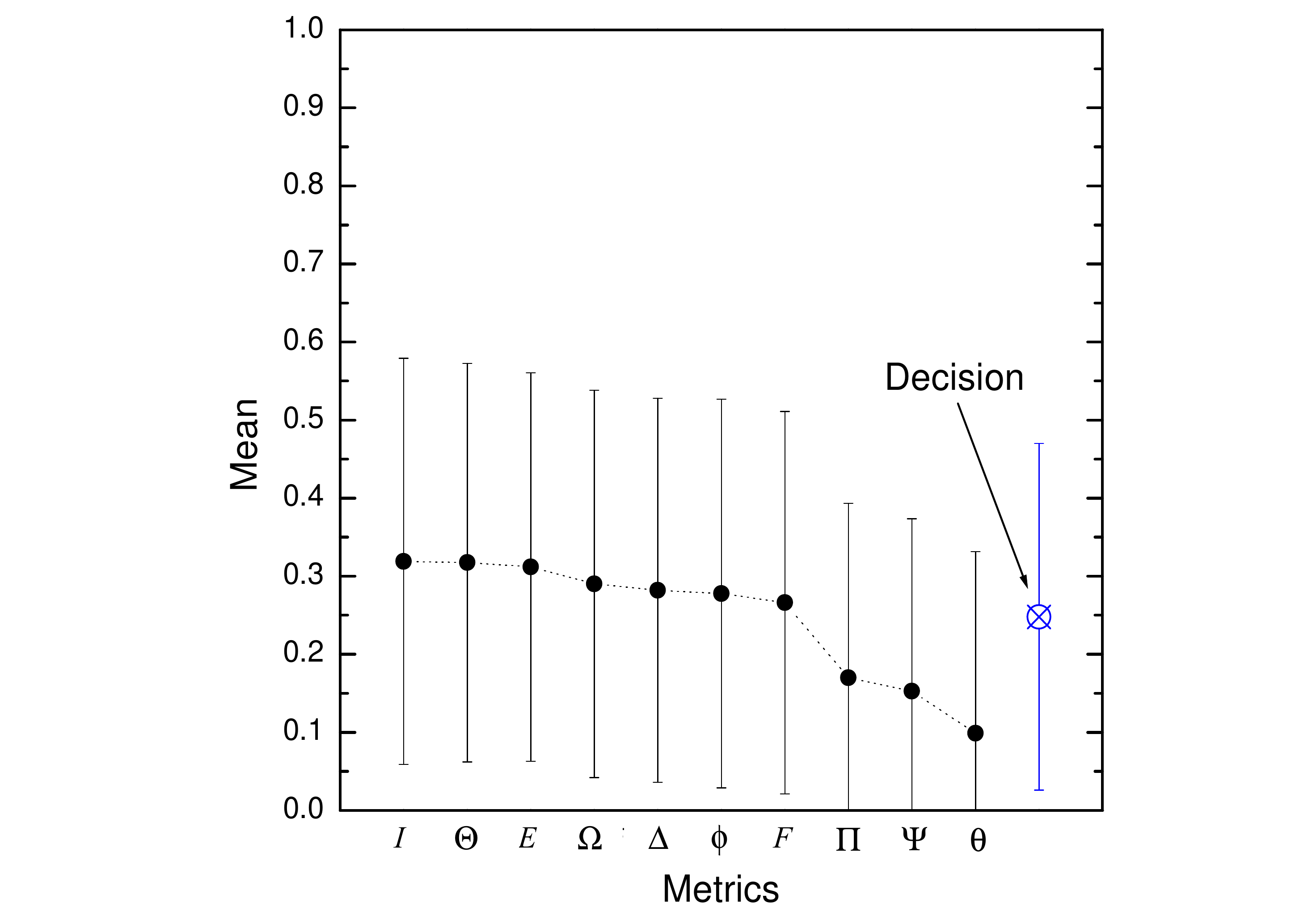}}
 \caption{\emph{\small On the left: standard deviation for each metrics. At the right:
          mean value of metrics and mean value of the Decision Algorithm.
          Tests over $19,090$ text documents that contain $916,170$ sentences have been performed.
          Metrics are: Frequency $F$, Interactivity $I$, Sum of Hamming weight of words per segment $\Theta$,
          Hamming distances $\Psi$, Hamming weight of segments $\phi$, Hamming weight heavy $\Pi$,
          Sum of Hamming weights of words by frequency $\Omega$, Entropy $E$, Sum of probability frequences $\Delta$ and Angle 	
          $\theta$.
         }}
 \label{fig:m-metricas}
\end{figure}

We have a set of metrics and a decision algorithm that give a score
for each sentence. However, the metrics are not equally important.
What about the metrics {\itshape capacity} to discriminate the
segments? Imagine the following situation: four metrics $\lambda_i;
i=1,2,3,4$ are applied to a document split into six segments: $s_j;
j=1,\cdots,6$. Metrics $\lambda_1$ gives the maximal value to all
segments, therefore its mean value $\left\langle
\lambda_1\right\rangle=1$ and its variance $\sigma_1=0$. Metrics
$\lambda_2$ rejects all segments, then
$\left\langle \lambda_2\right\rangle=0$ and $\sigma_2=0$. Metrics
$\lambda_3$ evaluates all  segments with {\slshape the
same} value $\left\langle \lambda_3\right\rangle=0.5$, then
$\sigma_3=0$. Finally, metrics $\lambda_4$ gives maximal value to three
segments and 0 to the rest: $\left\langle
\lambda_4\right\rangle=0.5$ and its variance $\sigma_4 \approx
0.547$. Which of these four metrics  is the best? The answer is
related to the metrics capacity to separate the pertinent
segments from the non pertinent ones. The mean value does not represent
this measure, since $\lambda_1$ and $\lambda_2$ have the "same"
constant value (always 0 or always 1). None of them is discriminant, yet they have extreme mean values. $\lambda_3$ is still worse,
because it is an undecided metrics: it is incapable to decide "yes"
or "no". Finally, metrics $\lambda_4$ has the same average as
$\lambda_3$ (0.5), but unlike the others, its variance is important.
This metrics is better in separating the segments. So, the variance may be be used to calculate the metrics sensibility.
Indeed we performed a statistical study to evaluate this capacity. We
calculate the sensibility values of $\approx$ 20,000 documents over
$\approx$~1 million sentences. The result is shown in Figure
\ref {fig:m-metricas}, on the left side. In this figure, it is clearly
visible that all the metrics have high sensibility values, then all
of them are important. This is a suitable property but, what about
the metrics mean value? In the right side, we plotted the mean value
of each metrics. In other words, this value represents the
average compression rate $\left\langle\rho_i\right\rangle$ we obtain
with the metrics $i$. Therefore, metrics angle
$\theta$, Hamming's distances $\Psi$ and Hamming weight heavy $\Pi$
eliminate, in the average, among 80\% and 90\% of the text's phrases. The
rest of the metrics eliminates 70\%. The Decision Algorithm, in the average,
retains close to 25\% of sentences.
This first study shows that all the metrics are discriminant
and that they have the ability to condensate text at high rates.

\begin{figure}
 \centerline{\includegraphics[width=14pc]{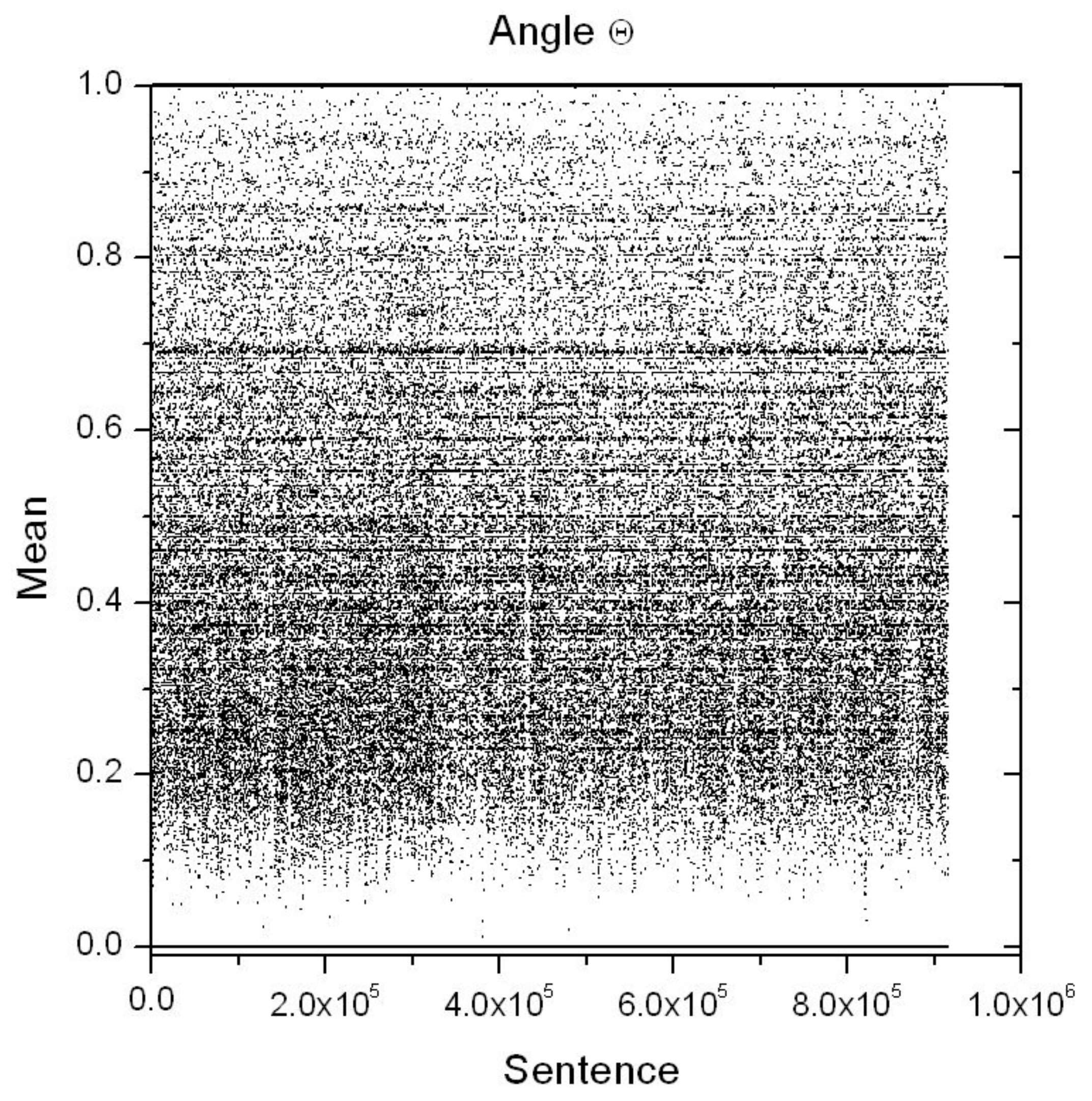}
             \includegraphics[width=14pc]{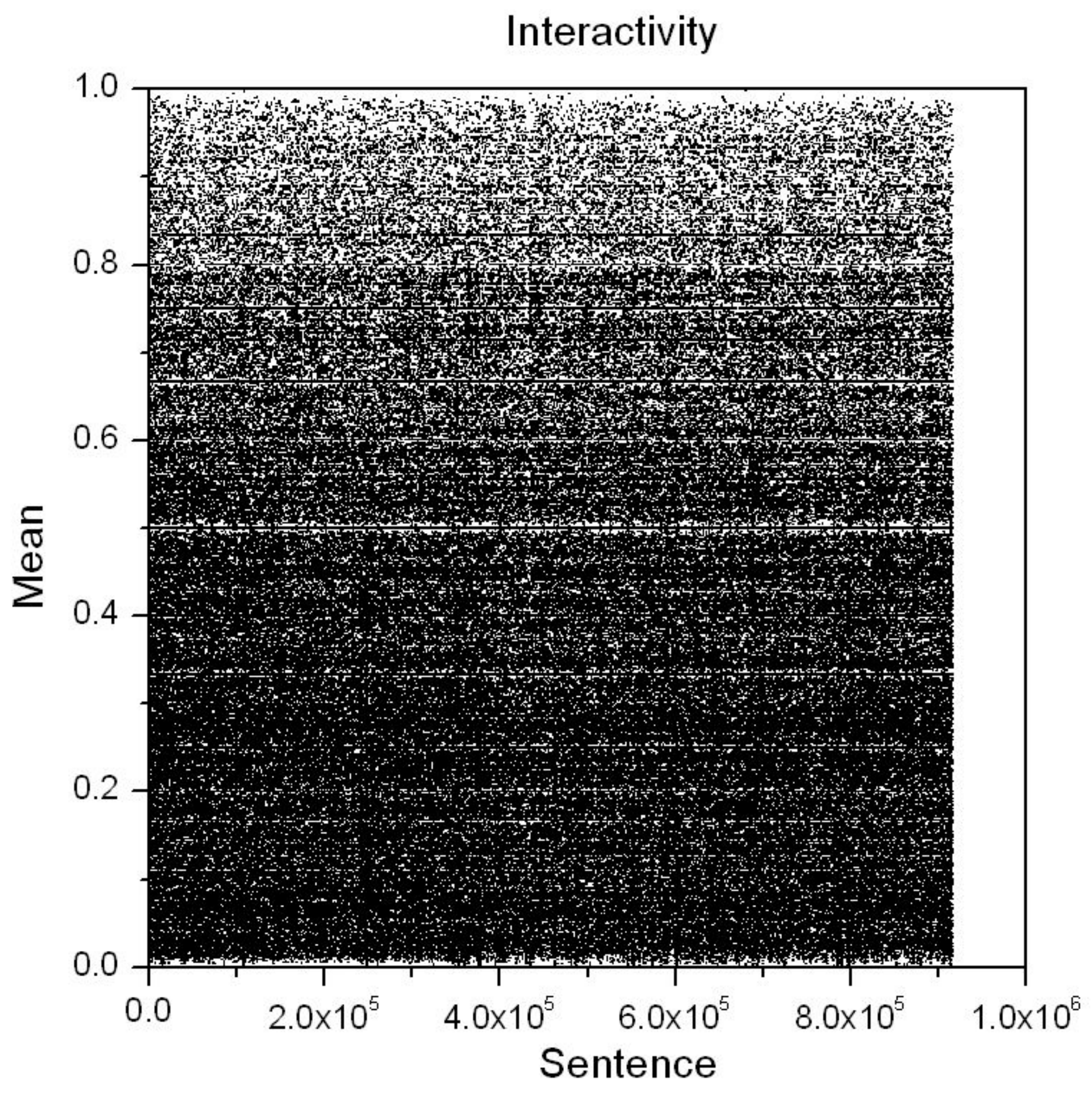}}
 \caption{\emph{\small Density means of decision "yes" (advantaged sentence) and "no" (disadvantaged sentence)
          on $916,170$ sentences, for Angle $\theta$ and Interactivity $I$ metrics. 
          Every point represents the normalized value $\lambda^{\mu}$
          calculated by the metrics on sentence $\mu$.
          On the left, we show the Angle metrics $\theta$. Most values for this metrics
          are on the bottom (i.e. the metrics decides 0.00, so it strongly disadvantages many sentences) and they are not visible
          because they are mapped to horizontal axis. Then a sparse density is found.
          On the right, the interactivity metrics $I$ is shown. Most values
          are under 0.5, but the $I$ density is more uniform than $\theta$, i.e. this metrics is less
          decisive than $\theta$ (see Figure \ref{fig:kappa} for more details).
 }}
 \label{fig:decision-metricas}
\end{figure}

\begin{figure}
 \centerline{\includegraphics[width=13.5pc]{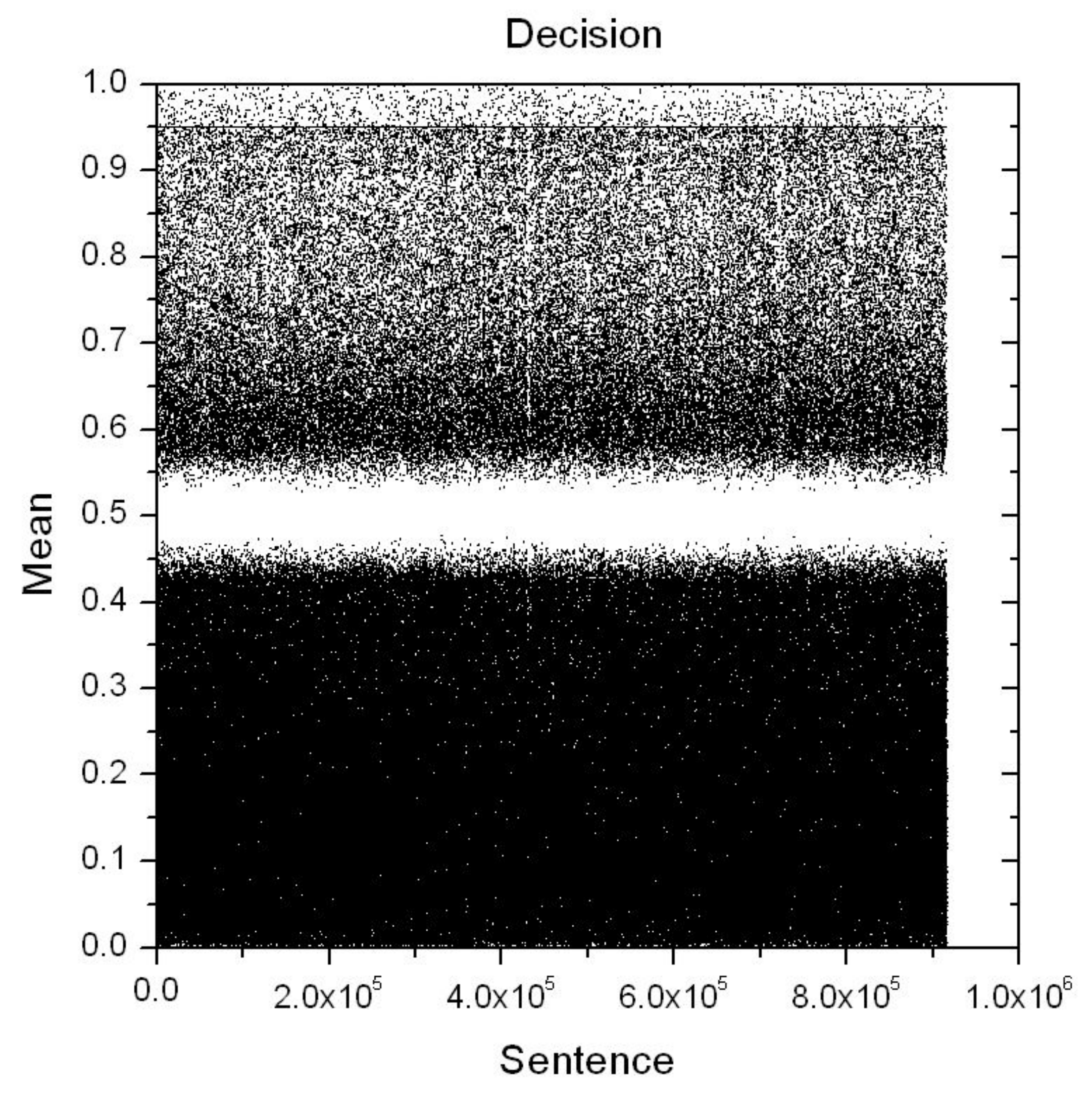}
             \includegraphics[width=18pc]{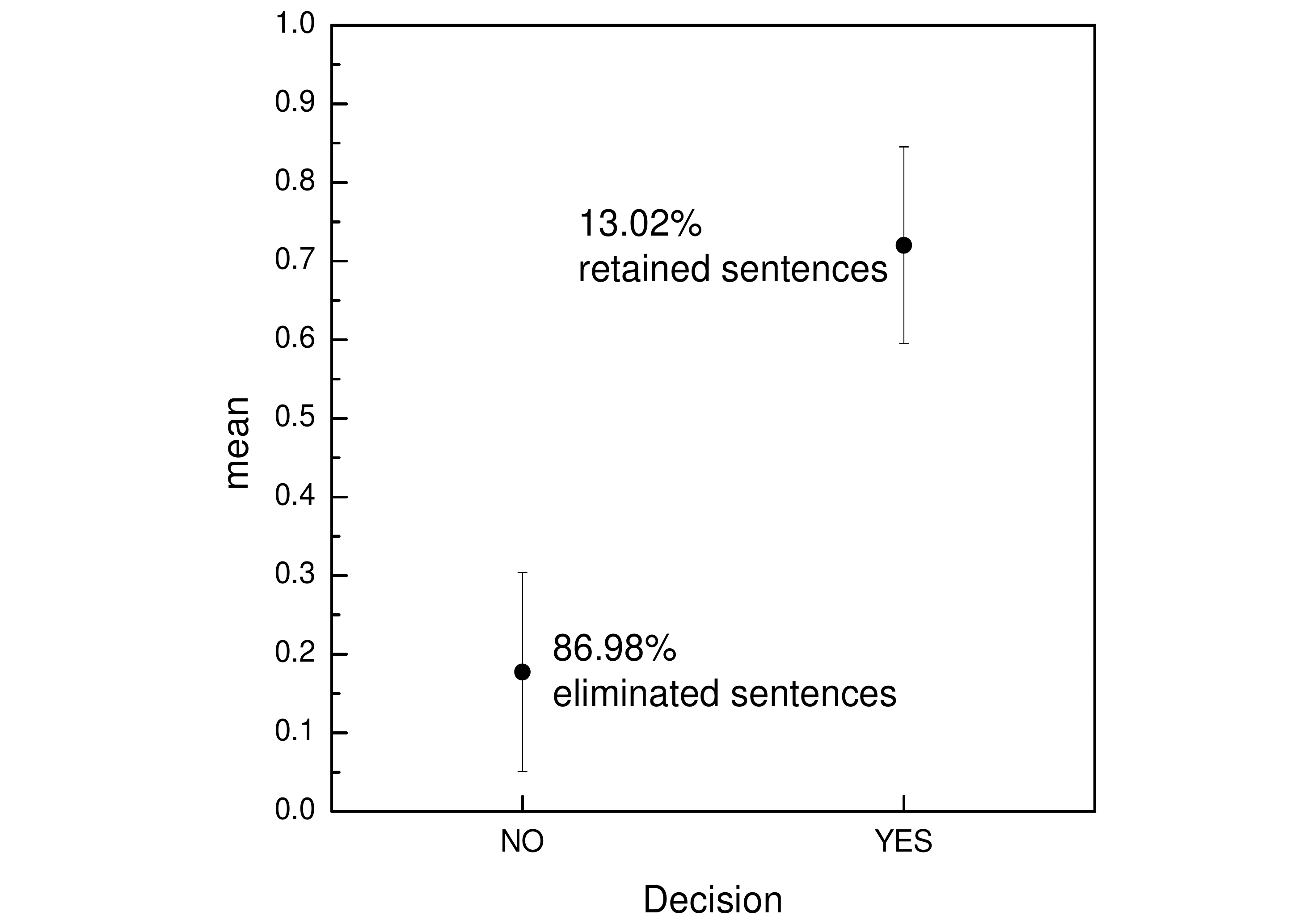}}
 \caption{\emph{\small On the left: "density" of the decision algorithm is plotted over $916,170$ sentences.
          On the right we show mean values of Decision Algorithm for "yes" (advantage sentence) and "no" (disadvantage sentence).
          It is visible that the Decision Algorithm is powerful (there are no "undecided" sentences, i.e. with $\Lambda=0.5$ in
          a gap of $\approx 0.1$), and most sentences (near 87\%) are disadvantaged.
          Effective compression rates $\left\langle\kappa\right\rangle^\pm$ are calculated with
          equations \ref{eq:rates1} and \ref{eq:rates2}.
 }}
 \label{fig:decision}
\end{figure}

\begin{figure}
 \centerline{\includegraphics[width=16pc]{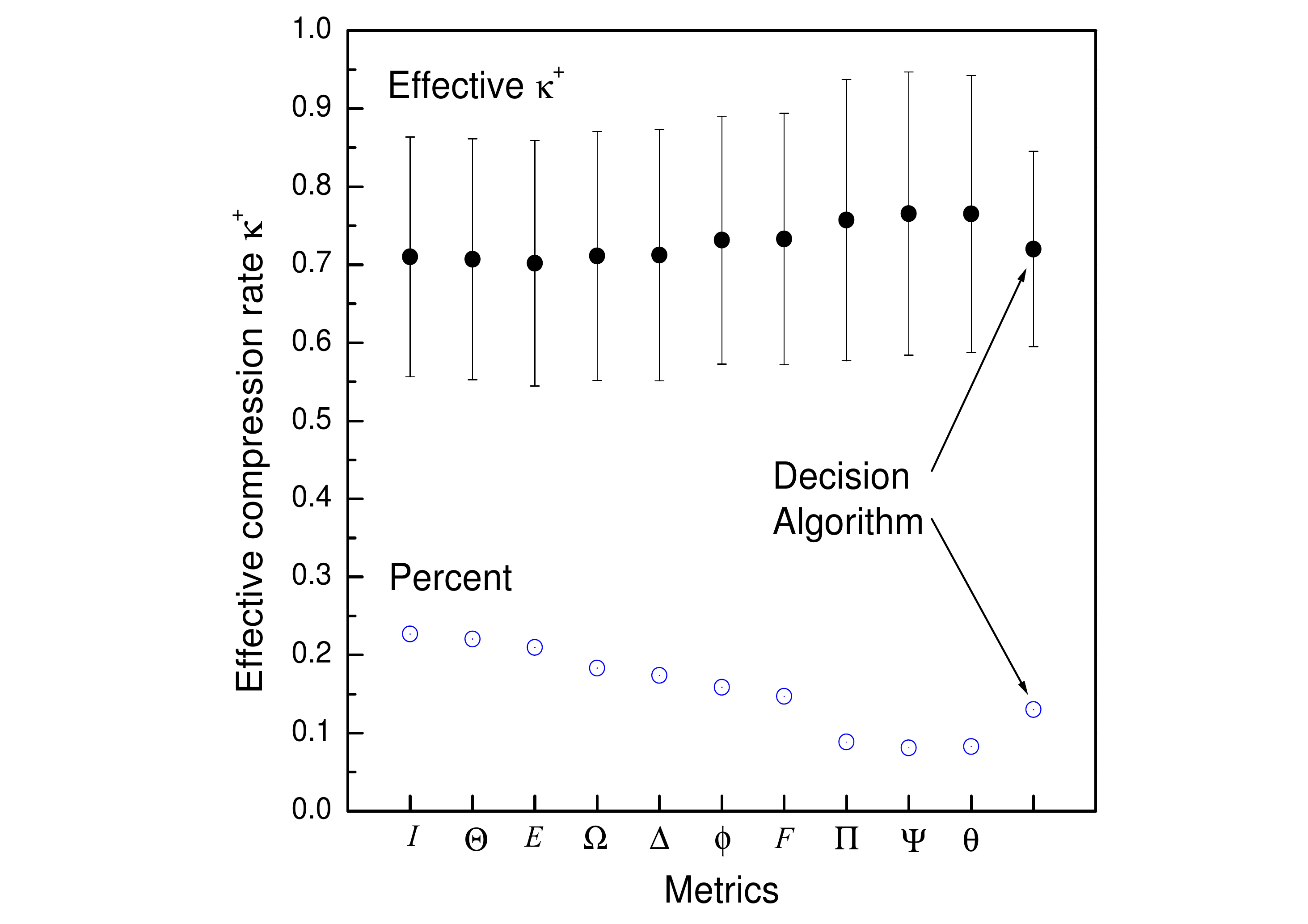}
             \includegraphics[width=16pc]{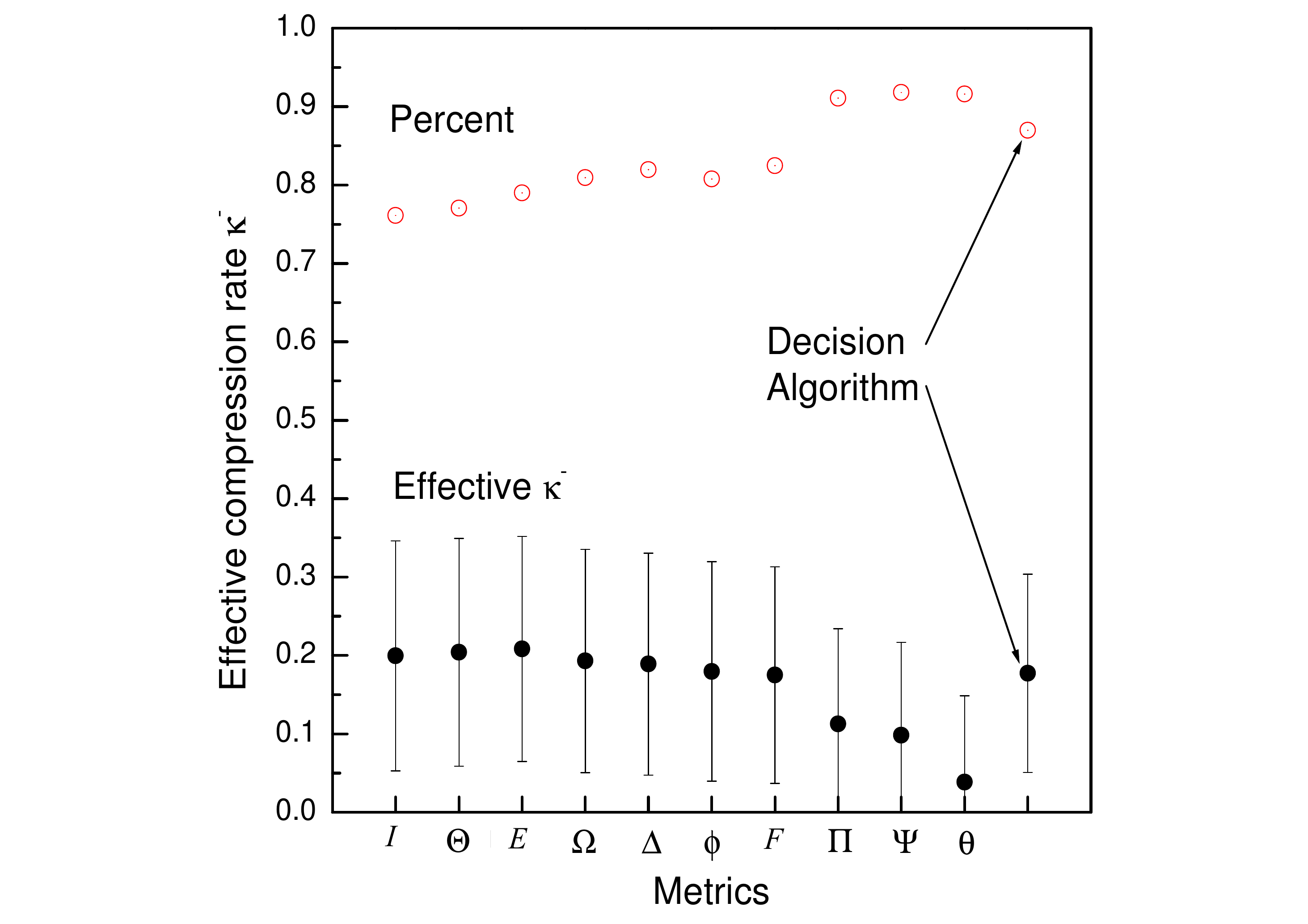}}
 \caption{\emph{\small Effective compression rate $\left\langle\kappa\right\rangle^\pm$ over $916,170$ sentences.
          On the left, we show $\left\langle\kappa\right\rangle^+$ and the percentage of sentences retained.
          At the right, $\left\langle\kappa\right\rangle^-$ and the percentage of sentences eliminated.
          The order of metrics on the horizontal axis is the same as that in Figure \ref{fig:m-metricas}.
          On the left, we note that the metrics Angle $\theta$ retains a low percentage of sentences ($\approx 10\%$)
          with high decision values ($>0.7$). The observation for the same metrics, in the right, shows that this metrics
          eliminates a big percentage ($\approx 90\%$) of sentences with very low values ($< 0.05$).
 }}
 \label{fig:kappa}
\end{figure}

However, a finer study of the metrics and Decision Algorithm has been performed.
We have considered the proportion of advantaged and disadvantaged segments separately.
In Figure \ref{fig:decision-metricas}, we show only two metrics, the first one representing
the density picture for angle $\theta$, and the second one, the density shape for interactivity $I$.
In Figure \ref{fig:decision}, on the left, we show the density picture for the Decision Algorithm.
It is clear that there are no undecided values ($\Lambda^\mu\neq\frac{1}{2}$) in the Decision Algorithm,
and that most  of the sentences ($\approx 87\%$) have been disadvantaged ($\Lambda^\mu<\frac{1}{2}$).
We defined the effective mean compression rates $\left\langle\kappa\right\rangle^+$ and $\left\langle\kappa\right\rangle^-$ for every metrics as follows:

\begin{eqnarray}
 \label{eq:rates1}
 \left\langle\kappa\right\rangle^+ & = &
              \frac{1}{\mbox{card}\{{\lambda^\mu > 0.5}\}} \sum^{N_S}_{\begin{subarray}{c}
                \mu = 1 \\
                \lambda^{\mu} > 0.5
                \end{subarray}} \lambda^\mu \\
 \label{eq:rates2}
 \left\langle\kappa\right\rangle^- & = &
              \frac{1}{\mbox{card}\{{\lambda^\mu < 0.5}\}} \sum^{N_S}_{\begin{subarray}{c}
                \mu = 1 \\
                \lambda^{\mu} < 0.5
                \end{subarray}} \lambda^\mu
\end{eqnarray}

\noindent where $\mbox{card}\{\bullet\}$ represents the cardinality of set $\{\bullet\}$. In Figure
\ref{fig:decision}, on the right, values $\left\langle\kappa\right\rangle^\pm$ are shown for every metrics and for the DA. In Figure \ref{fig:kappa}, the effective compression rate $\left\langle\kappa\right\rangle^{\pm}$ and its
corresponding ratio (in percent) of advantaged  or disadvantaged sentences is shown.

\subsubsection{Order of presentation of segments}

Another important robustness test was performed. We mixed the sentences at random to generate a new text.
This text was then processed by to {\sc Cortex} and the same values for the Decision Algorithm were found.
Our results showed that the order of presentation of sentences
has no impact on the final decision of the DA.
This can be explained because our algorithm does not use any position-sentence metrics.
On the other hand, tests on Minds and Word summarizers show that these methods
are sometimes dependent on the order of presentation of sentences.
Indeed, the segmentation of sentences by the separator $\left\langle:\right\rangle$ tends to perturb
their performance.

\section{Evaluation}

The best way to evaluate automatic text summarization systems is not evident, and it is still an open problem \cite{FIR99}. 
In general, methods for evaluating text summarization systems can be classified into two main categories \cite{IND99}. 
One is an intrinsic evaluation, where humans judge the summary quality directly. 
However, this approach is very difficult to implement in the case of big corpora (for example, if a multi-document corpus must be summarized). 
Therefore extrinsic methods will be necessary in this situation.

In an extrinsic evaluation, the summary quality is judged based on
how it affects the performance of other tasks. We
choose the coupling with a Question-Answering system to perform this
evaluation. Formally, for extrinsic evaluation, we applied
Confidence-Weighted Score ($CWS$) \cite{VOR02} to evaluate
the output of the QA system. $CWS$ was specifically chosen from
TREC-2002 to test a system's ability to recognize when it has found
a correct answer. The questions were ordered in such a way that the highest in ranking
was the question for which the system was most confident in its
response and the lowest was the question for which the system
was least confident in its response. 
If two or more systems produce the same set of candidate answers, but in a different ordering, the system which assigns the highest ranking to the correct answer is regarded as the best one.
Formally the confidence-weighted score is defined as:

\begin{equation}
  CWS = \frac{1}{Q} \sum_{i=1}^Q \frac{i_c}{i}
  \label{eq:CWS}
\end{equation}

\noindent  $Q$ is the number of questions and $i_c$ the number of correct answers in the first $i$ questions
(position within the ordered list). The $CWS$ criterium is used to order the selected candidate answers 
according to the score of the sentences provided by the personalized {\sc Cortex} system in Extrinsic methods (see Section \ref{Sec:exp2}).

\section{Experimental framework I. Intrinsic methods: generic abstracts}

In \cite{TOR04,TOR01,TOR02} we showed results of the {\sc Cortex} system applied to generic summaries. Its performance is
better or equal than that of other generic summarization methods. We will reproduce some results of these experiments here.
We tested the algorithm on documents coming from various sources (scientific articles, excerpts of the press on the Internet). We compared our results with Minds\,\footnote{http://messene.nmsu.edu/minds/SummarizerDemoMa in.html}, Copernic summarizer\footnote{http://www.copernic.com}, Pertinence\footnote{http://www.pertinence.net}, Quirck\footnote{http://www.mcs.surrey.ac.uk/SystemQ/}, Word and
baseline systems. In order to evaluate the quality of summaries, we compared all the results 
with summaries produced by 17 people (students and university professors) accustomed to write summaries.

Some tests on the text "Puces" (see Annexe \ref{ann:puces}), which is artificially ambiguous (because of its heterogeneous mixture of texts from two different
authors) will be presented. 
The subject "computer chips" is in the first part ($\approx 2/3$) of the text, and the presence of fleas in a Swiss military
company\footnote{http://www.admin.ch/cp/f/1997Sep10.064053.8237
  @idz.bfi.admin.ch.html} is discussed in the second one\footnote{In
  French, the word "{\itshape puces}" (fleas/chips) is ambiguous in
  this context.}. 
Obviously, no hints of this preliminary knowledge are submitted to the system. 
This text contains $N_W=605$ words. 
The segmentation process splits the text into $N_S=30$ sentences. 
Then, filtering/lemmatization/stemming process returns a set of $N_M=279$ terms. 
It contains $N_L=30$ distinct terms. 
The topic of sentences 0 to 14 is about computers chips, whereas sentences 15 to 29 discuss fleas. 
An abstract of 25\% of the original text size must contain 8 sentences. 
We expected that the systems would produce a summary composed of two sub-summaries (taking into account both subjects).
This result was well confirmed for our algorithm.

\begin{figure}
 \centerline{\includegraphics[width=18pc]{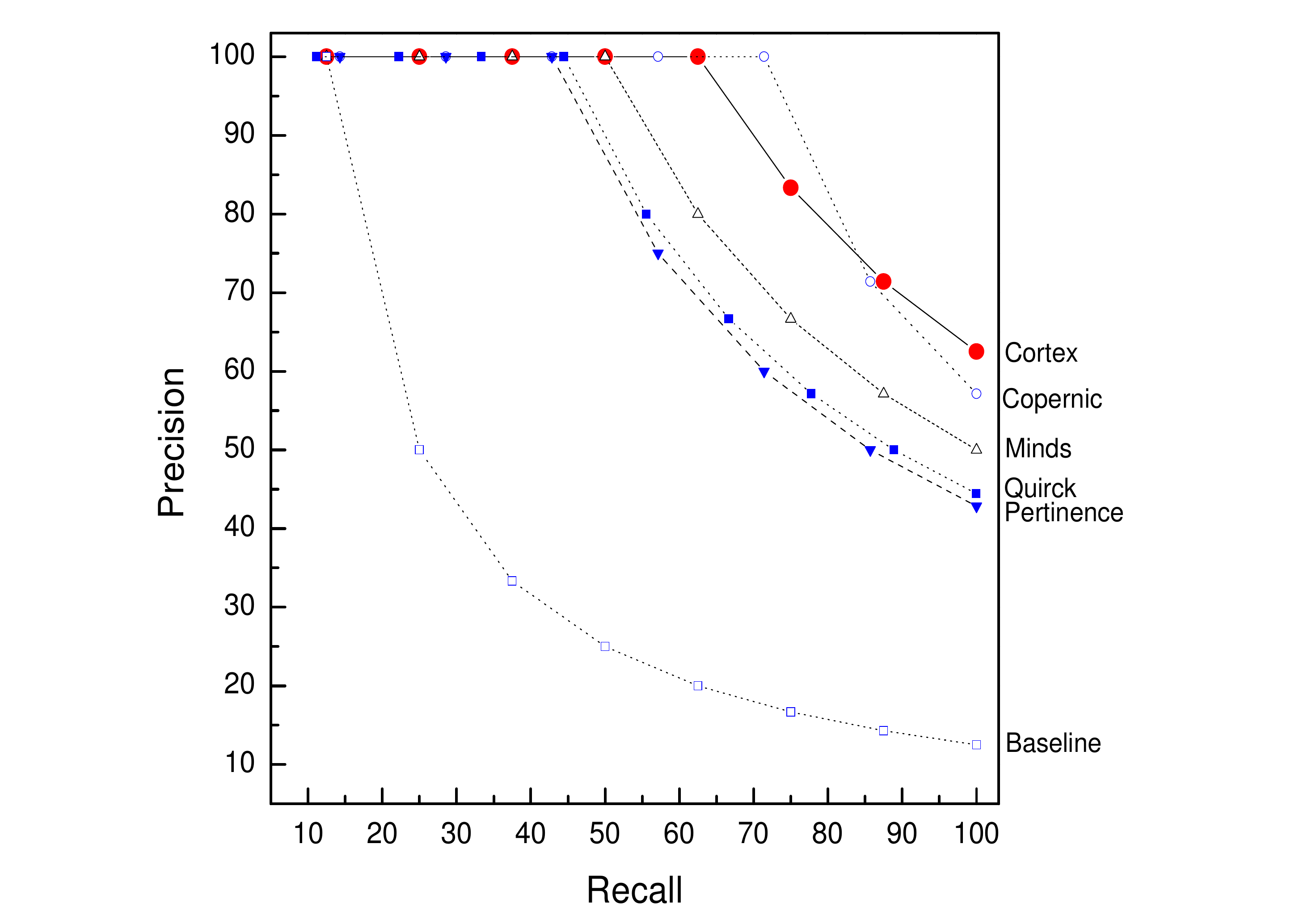}}
 \caption{\emph{Precision-Recall graphic for several algorithms on text "Puces".
          At 60\% recall, only {\sc Cortex} and Copernic yield 100\% precision.
          At 100\% recall, {\sc Cortex} shows 62.5\% precision.
          We do not show values for the Word summarizer because precision and recall are
          both 0 in this task.}}
 \label{fig:puces-pre-rec}
\end{figure}

Figure \ref{fig:puces-pre-rec} shows the precision-recall plot for the "Puces" text.
In this graphic, the Copernic and {\sc Cortex} algorithms yield the best precision values for this task. 
{\sc Cortex} has a value of  62.5\% for precision at 100\% recall.
However, we think that the precision measure may be not sufficient 
to evaluate the quality of extracts. So, the Precision-Recall plot may be completed with an other
evaluation measure: the quality $\mathcal{Q}$. We evaluated the quality of extracts obtained for each method by
measuring the value:

\begin{equation}
 \mathcal{Q} = \sum^S_{\begin{subarray}{c}
             \mu \in \textrm{ extract}
             \end{subarray}} \mathcal{H}^{\mu} \Theta, \mbox{where~} \Theta = \left\{\begin{array}{l} 1 \textrm{ if segment } \mu \textrm{ is present in the human's extract}
                                                \\ 0 \textrm{ otherwise}
             \end{array}
             \right.
 \label{eq:quality}
\end{equation}

\begin{figure}
 \centerline{\includegraphics[width=16pc]{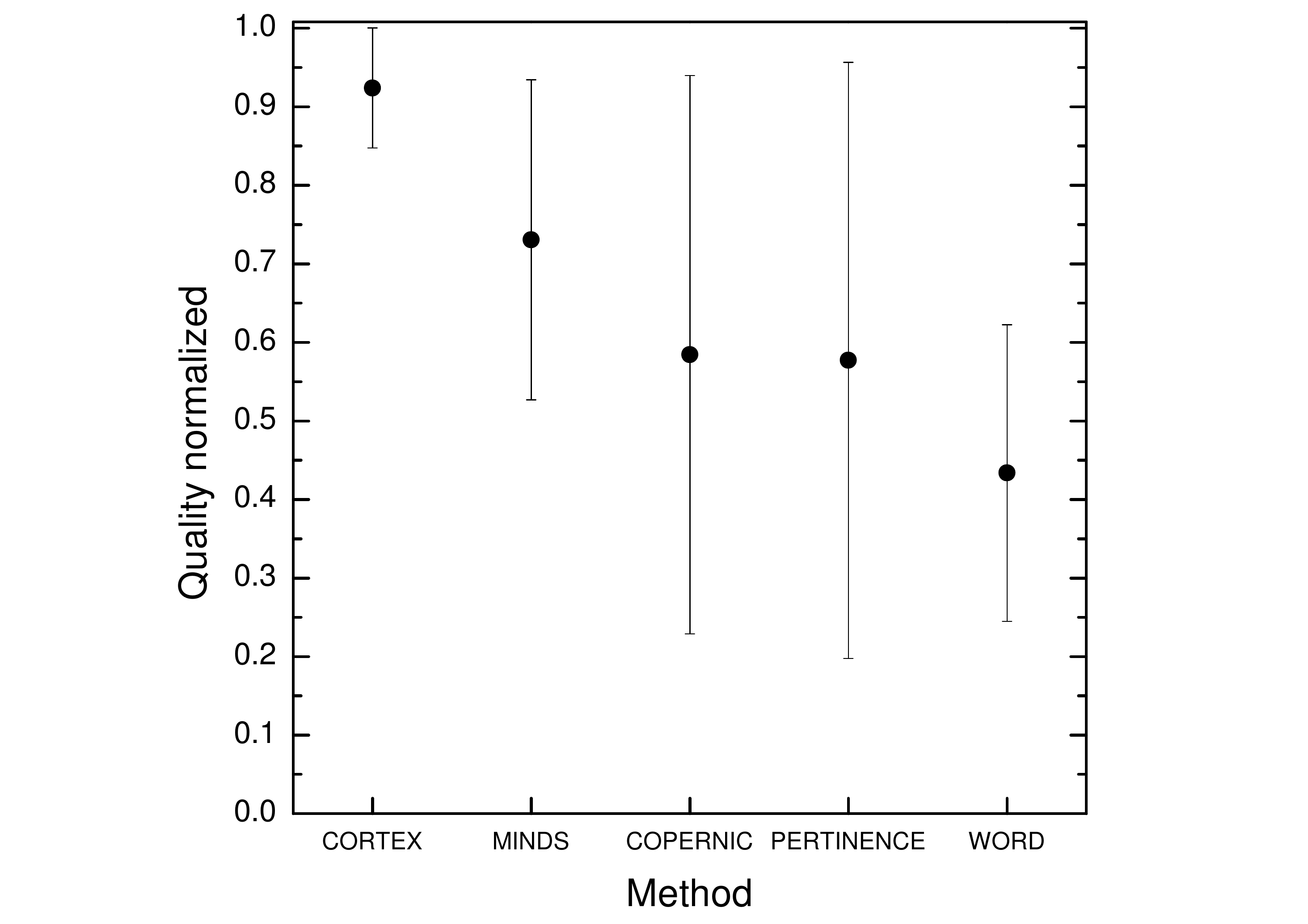}}
 \caption{\emph{Measuring the quality of methods.
         The vertical axis measures the normalized quality $\mathcal{Q}$ (calculated by equation \ref{eq:quality}) 
         of extracts over seven French documents.
         Our algorithm {\sc Cortex} performs better than other methods in this generic summarizing task.}}
 \label{fig:quality}
\end{figure}

\noindent $\mathcal{H}^{\mu}$ is the mean value for segment ${\mu}$ in the extract compiled by human judges. 
Thus, a normalized version of $\mathcal{Q}$ is a kind of precision of a method.
Values of normalized $\mathcal{Q}$ for the studied methods over seven French texts \cite{TOR01,TOR02} are plotted in Figure \ref{fig:quality}.
In this graphic, {\sc Cortex} obtains the best quality values.
Other results \cite{TOR01,TOR02} show that our system is noise robust,
less sensitive to the order in with the sentences are presented and
the summaries are balanced and mostly of a better quality.

\section{Experimental framework II. Extrinsic methods: coupling with the LIA-QA system}
\label{Sec:exp2}

We found it interesting to couple {\sc Cortex} system with the vectorial search engine LIA-SIAC \cite{BEL01} and LIA-QA system \cite{BEL03} to evaluate the impact on the answers' precision. 
This might be a way to measure the quality of summaries and
possibly to improve the QA system. Thus, two types of experiments were performed. In the first one, generic abstracts were generated from a multi-document corpus. The second experiment uses a modification of the {\sc Cortex} system to generate personalized abstracts. In both cases, when the digests are obtained, we used the QA system to find exact answers to specific questions.
The statistics were estimated over 308 questions which were automatically assigned a type and an associated Named Entity (NE). We used the {\sc Cortex} system as a noise eliminator, with a compression rate $\tau$=\{80\%,90\%\}, as a true abstracting system with a compression rate $\tau$=\{10\%, 20\%\} and finally as an intermediate system with a compression rate between $30\% \leq \tau \leq 70\%$. Comparisons with baseline and random systems will also be presented.

\subsection{Corpus description}

The whole corpus contains $D \approx 20,000$ articles from the French
newspaper \textsl{Le Monde}\footnote{www.lemonde.fr}, was used in the
evaluation Technolangue EVALDA/EQUER project\footnote{http://www.technolangue.net/article61.html}. 
It covers the years 1987 to 2002 and it is a highly heterogeneous proprietary corpus. This corpus contains text coded in
ISO-Latin, which we transformed into UTF-8 code before processing. A
set of $Q=308$ questions in French has been provided by the
company Sinequa\footnote{www.sinequa.com}. They correspond to the
translation of some questions used in TREC programs (a translation
of 1893 questions TREC may be found on RALI
site\footnote{www-rali.iro.umontreal.ca/LUB/qabilingue.en.html}).
Four examples of questions with their expected NE has been reproduced here:
\begin{itemize}
  \item \textit{A quelle distance de la Terre se situe la Lune ?} (What is the distance from the Earth to the Moon?) NE: <DISTANCE>
  \item \textit{En France, à combien s'élève le pourcentage de la TVA
  sur les Compact Disc ?} (What is the percentage of VAT on Compact Discs in France?) NE: <VALUE>
  \item \textit{Où la Roumanie se situe-t-elle ?} (Where is Romania located?) NE: <PLACE>
  \item \textit{Que signifie RATP ?} (What does RATP mean?) NE: <SOCIETY>
\end{itemize}

In addition to the set of questions, Sinequa provided us for
each question with a list of documents containing at least a common word
with the question, a labeled version of these documents. In this list
the types of named entities (nouns, dates, names of companies, places,
durations, sizes,...) are marked, moreover the type of required entity, if it exists.

\subsection{Selection of candidate answers}

The LIA-QA system approach for the selection of candidate answers is
used with the vectorial search engine LIA-SIAC \cite{BEL01} to 
localize the required entity type and in the exploration of
knowledge bases \cite{BEL03,GIL03}\footnote{It should be noted that
for the set of experiments presented here, we did not
exploit the knowledge bases usually employed by the LIA-QA
system.}. Initially, the documents, summaries or not, are split
into lexical units and sentences, labelled syntactically and then
lemmatized. 
After calculating the proximities of the sentences to the question they are ordered according to this values
(we use a weighting of the type
tf.idf and a cosine value) \footnote{In these experiments we have neither enriched nor modified the question, except
for the elimination of the function words and lemmatization.}. Then, a filtering can be applied to preserve only the sentences containing at least a named entity corresponding to the expected type for the question. In a simplistic way, the response of the system could be the first entity of expected type appearing in the sentence\footnote{This way of proceeding has at least two problems shortcomings: in filtering, while avoiding the sentences in a too draconian way, returns sometimes impossible the extraction of an answer and the selection of the "first" entity may lead to an inadequate choice in the case that several entities of the same type are present in the same sentence.}.

\subsection{Search of answers in generic summaries}

Tests on generic summaries were realized to verify their
capacity to preserve informative textual zones, that is, the zones suitable to answer to precise questions.
Tables \ref{tab:REF} and \ref{tab:EN} show that compression rate is not proportional to the number of correct answers returned by the system. 

{\sc Cortex} was applied on corpus to generate summaries at different compression rates $\tau=\{10\%, 20\%, 30\%,\cdots,90\%\}$. 
In this stage, all the $\Gamma=10$ metrics are used. 
Table \ref{tab:REF} shows the real rates observed after creation of the summaries according to the number of sentences, words or characters for some compression rates $\tau$.
\begin{table}[h]
\centerline{
\begin{tabular}{|c|c|c|c|}
\hline
  \textbf{Expected rate $\tau$ \%} & \multicolumn{3}{c|}{\textbf{Observed rate \%}}  \\
  \cline{2-4}
                & Sentences  &  Words & Characters\\
  \hline     \hline
  80  & 71.1 & 77.6 & 77.5 \\
  \hline
  50   & 45.8 & 59.3 & 59.2 \\
  \hline
  20  & 20.6 & 31.5 & 31.5 \\
  \hline
\hline
\end{tabular}
}
 \caption{\emph{Real compression rate of summaries in number of sentences, words and characters, in function of desired rate.}}
 \label{tab:REF}
\end{table}


Table \ref{tab:EN} depicts the results obtained when the QA system was coupled to the generic {\sc Cortex} system (generic QAAS).
We expected that if the QA system worked on summaries instead of  full text, it would reject  non-informative areas that may generate a false hypothesis, and preserve those which will provide right answers  to the questions. In addition, we also hoped to save time because of the reduction of the search space.
The score of each sentence was calculated by the LIA-SIAC engine, after wich the LIA-QA system uses this score to 
find the answers.


\begin{table}[h]
\centerline{
 \begin{tabular}[t]{|l||c||c|c|c|c|c|c|c|c|c|c|}
 \hline
  \textbf{Compression }     & 100   &      90    & 80    & 70   & 60  & 50  & 40 &   30 & 20   &  10  \\
  \textbf{Rate $\tau$ \%}   &       &            &       &      &     &     &    &      &      &      \\
 \hline\hline
   Responses                & 187   &      187   & 187   & 187  & 187 & 187 & 187& 187  & 185  &  179 \\
 \hline
   Correct answers          & 50    &      46    &  46   & 42   & 42  &  42 &  38& 38   &  33  &  26  \\
 \hline
  $CWS$                     &{\bf 30.89}& 25.33&{\bf 25.34}&25.33&24.46&24.32&23.03&22.11&21.78&17.38\\
 \hline
 \end{tabular}
 }
 \caption{\emph{Correct answers, responses and $CWS$ (see equation \ref{eq:CWS}) found by the generic {\sc Cortex} system.
          We show that the number of responses and correct answers
          is slightly degraded by high compression rates ($\tau$<50\%).}}
 \label{tab:EN}
\end{table}


\subsection{Search of answers in user's query-based summaries}
\label{Sect:personn}

We have demonstrated how coupling a question-answering system with  a text summarization system may make the latter one more efficient by reducing the search space, without significantly altering  the quality of results. 
However, the use of generic summaries might be
limited. One may hope that a query-oriented summary could find the 
answers more efficiently, because the documents would be condensed in a targeted
way. In this section, we explain how to adapt the generic summarization
system to obtain a customized summarization system, whose behaviour
is adapted to the questions submitted by the user. The
personalization of summaries (taking into account  the user's
question) would increase the chances of not eliminating correct
answers. We have good reasons to think that this will improve the
precision of the answers. Figure \ref{fig:lia-qaas2} shows the
architecture of the LIA-QAAS (Question-Answering Automatic-Summarization) system. 

First, LIA-SIAC extracts a subset of $R_D$ relevant documents for each question from the corpus. Concurrently, the set
of $Q=308$ questions is filtered, lemmatized and stemmed. An expansion process (described in the next section) is applied to this question set. Thereafter a multi-document abstract at variable compression rates ($10\% \le \tau \le 90\%$) is obtained by {\sc Cortex}. 
In this stage, the score for each sentence is local to each document. In the next step, 
the multi-document abstract are re-scored by using the {\sc Cortex} system once more.
The result consists in a set of query-personalized sentences sorted by score for each question. 
The process will be described in the next subsection.


\subsubsection{Adding search terms to a user's query}

Query expansion consists of adding search
terms to a user's weighted search. For example, a search for {\bfseries car}
may be expanded into \textbf{\{car, cars, auto, autos, automobile, automobiles\}} and then lemmatized to
\textbf{\{car, auto, automobile\}}. Query expansion  was applied to this set by adding synonymous 
terms taken from a simple thesaurus. This will result in one
or two additional terms for each term in the user's query. Query expansion has
the disadvantage that undesirable noise may be added, but the purpose
is to improve precision and/or recall by using a more flexible query.
In our case the introduction of noise is minimized, because the
expansion is applied only to the query and not to full text. 

Formally each query is represented by a vector $\vec q_j,
\mbox{where}~ j=1,\ldots,Q$. For each document in the corpus, its main
title, represented by vector $\gamma^T$, is substituted by the set of
every vector query $\vec q_j$ such that its answer is likely to be
found in the document (see equation \ref{eq:theta}). 
The metrics used are only Frequency $F$ and Angle $\theta$. 
This combination measures exactly the similarity between $\vec q_j$ and each sentence
$\xi^{\mu}_j$, i.e. the sentences that are closer to user's
question. 

\begin{figure}
 \centerline{\includegraphics[width=21pc]{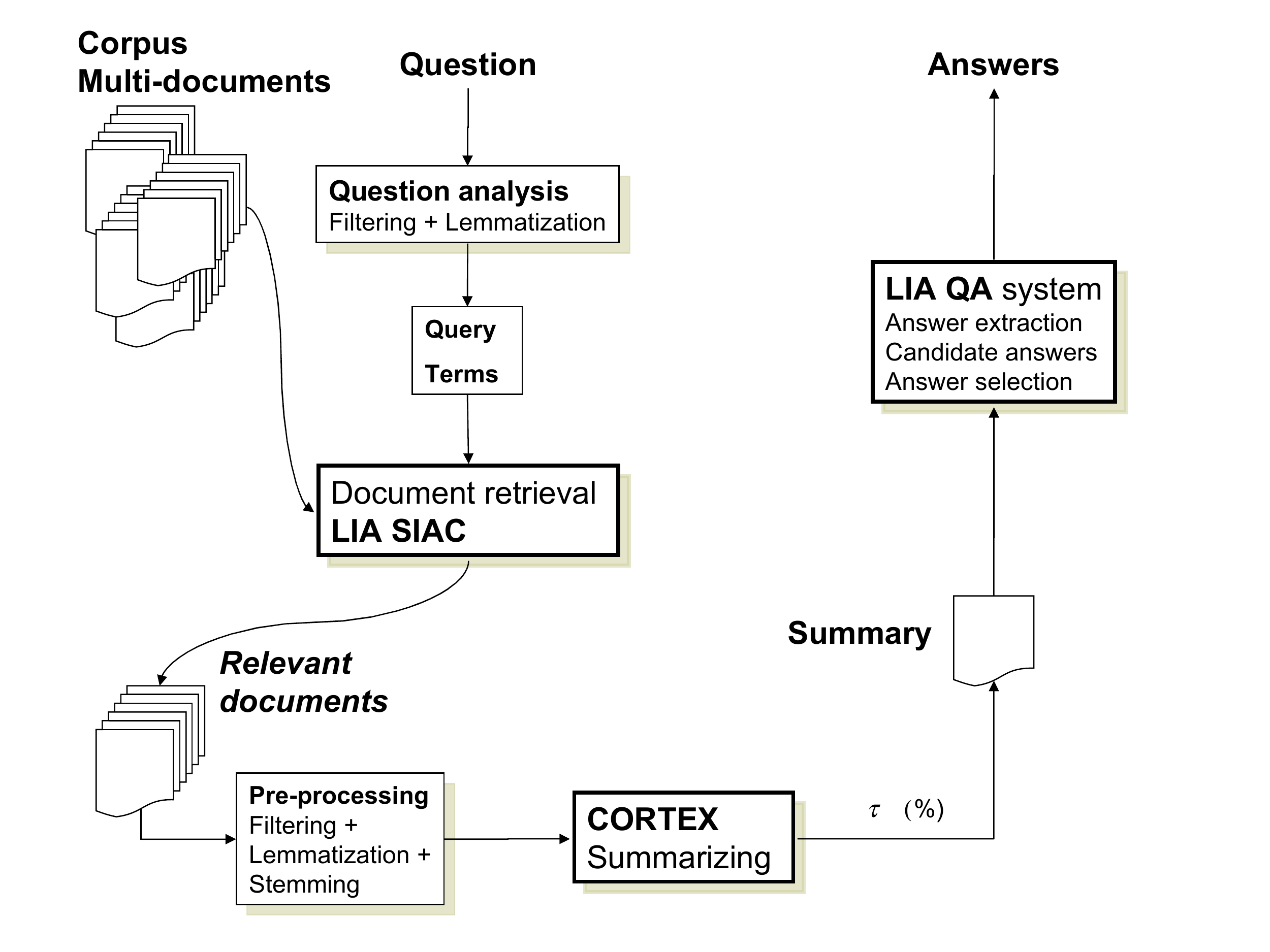}}
 \caption{\emph{The LIA Generic Question Answering-Automatic Summarization (QAAS) system.
          A user's natural language question is transformed into a query. 
          Candidate documents are chosen from multi-document corpus by LIA-SIAC, and then pre-processed.
          {\sc Cortex} summarizes this multi-document subset to generate a generic summary. 
          The LIA-QA system is applied to this summary to generate an answer to the question.}}
  \label{fig:lia-qaas-gen}
\end{figure}

\begin{figure}
 \centerline{\includegraphics[width=21pc]{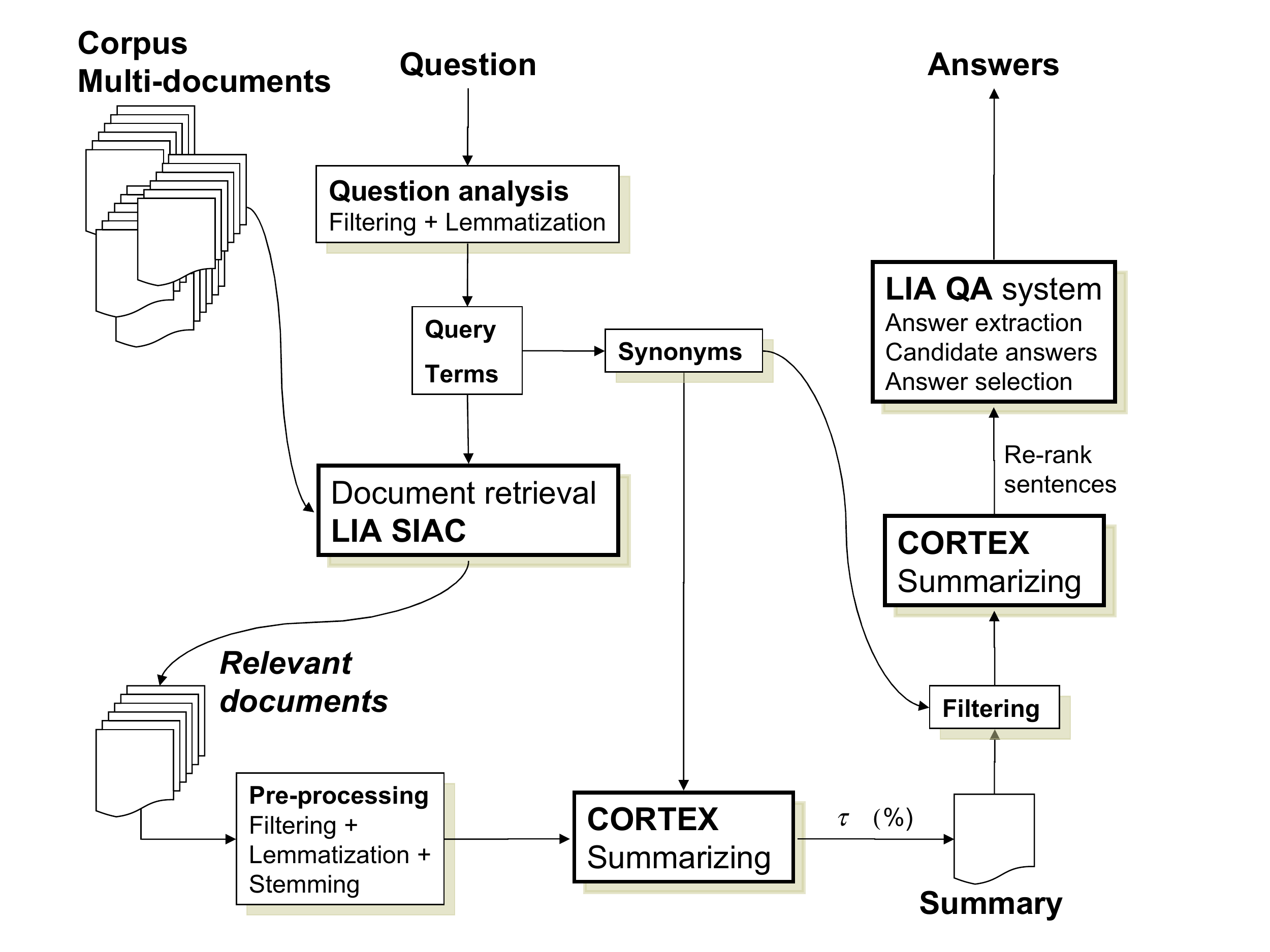}}
 \caption{\emph{The LIA Personalized QAAS system.
          A user's natural language question is transformed into a query. Then query expansion is applied.
          Candidate documents are chosen from multi-document corpus by LIA-SIAC, and then pre-processed.
          {\sc Cortex} summarizes this multi-document subset to generate a customized summary. This summary  is
          filtered and sentences are re-ranked again by {\sc Cortex}. Finally the LIA-QA system is applied to 
          generate an answer to the question.}}
  \label{fig:lia-qaas2}
\end{figure}

Then, a set of abstracts at variable compression rates $\tau$ is generated by the {\sc Cortex} system. At the end of this process, we
obtain the most informative text areas for each document that match
with the query. Finally, a multi-document abstract is generated for
each question. In this stage, each sentence is locally ranked (the
sentences came from a particular text, then ranked with sentences from
the same text). A picture of the LIA-QAAS system (generic and personalized) 
is shown in Figures \ref{fig:lia-qaas-gen} and \ref{fig:lia-qaas2} respectively.

\subsubsection{Re-ranking of candidate segments}

At this stage, a summary has been generated for each question in the
multi-document corpus. Since each sentence's score is local  to
one document, several sentences may have the same score (for exemple,
many sentences may have a local score decision $\lambda=1.0$, and 
must be globally re-scored to avoid decision conflicts in QA system).
Another re-ranking process is applied to obtain a unique global score
(that takes into account all the documents for the query) per sentence. This
process returns a global score for each sentence that
depends on degree of similarity of the query. In this phase, terms in the document that are not present in the
vector query are filtered out. We obtain a new set of documents to which {\sc Cortex} is re-applied with all metrics.
In table \ref{tab:EN-custom} we show the results found by the QA system coupled to the query guided {\sc Cortex} system (personalized QAAS).

\begin{table}[h]
\centerline{
 \begin{tabular}[t]{|l||c||c|c|c|c|c|c|c|c|c|c|}
 \hline
  \textbf{Compression }     & 100   &      90    & 80    & 70   & 60  & 50  & 40 &   30 & 20   &  10  \\
  \textbf{Rate $\tau$ \%}   &       &            &       &      &     &     &    &      &      &      \\
 \hline\hline
   Responses                & 187   &      187   & 186   & 186  & 186 & 186 & 186& 186  & 186  &  186 \\
 \hline
   Correct answers          & 50    &      52    &  54   & 52   & 53  &  51 &  53& 54   &  52  &  52  \\
 \hline
  $CWS$                     &{\bf 30.89}&  29.69 & 29.96 &28.97 &29.69&30.00& 29.30&{\bf 30.75}&30.13& 29.66\\
 \hline
 \end{tabular}
 }
 \caption{\emph{Correct answers, responses and $CWS$ (see equation \ref{eq:CWS}) generated by the personalized {\sc Cortex} system
          (QAAS). We show that the number of responses and correct answers
          is not at all correlated to the compression rates. At compression $\tau=30$\% we found four more 
          correct answers (54) than when analysing full text (50) and $CWS$ is very close (30.75 vs. 30.89).}}
 \label{tab:EN-custom}
\end{table}

\begin{figure}
 \centerline{\includegraphics[width=22pc]{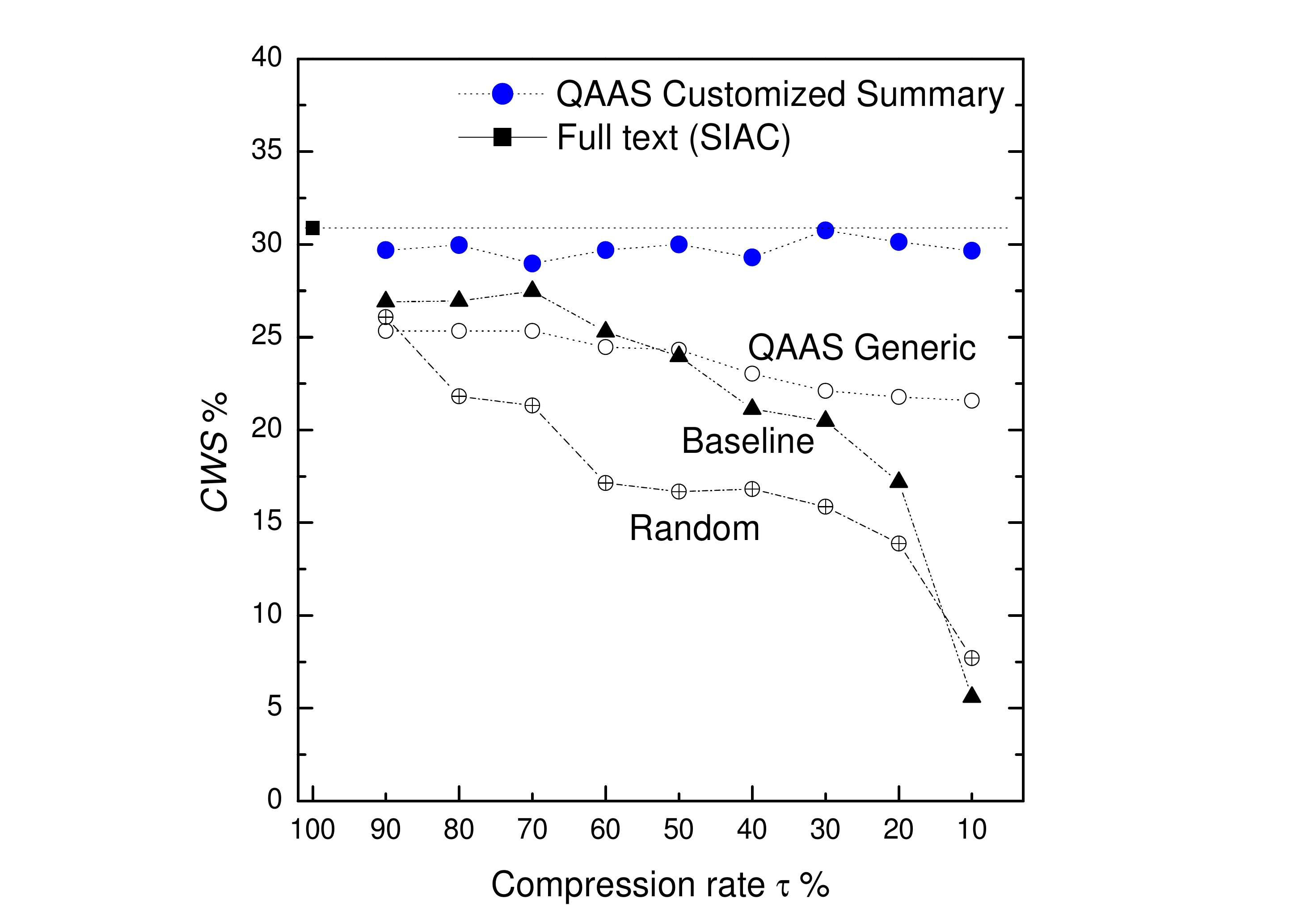}}
 \caption{\emph{Values $CWS$ for customized vs. generic abstracts, random and baseline systems. 
          The best personalized QAAS value $CWS = 30.75$ (full cercle pattern) for
          $\tau=30\%$ is very close to the value $CWS = 30.89$ obtained with full text (full square pattern). 
          In the case of generic abstracts, $CWS$ values are lower.
          This shows that query-based summaries filters and preserves
          the most informative segments of each document.
          }}
  \label{fig:CWS-PERS-GEN}
\end{figure}

\begin{figure}
 \centerline{\includegraphics[width=18pc]{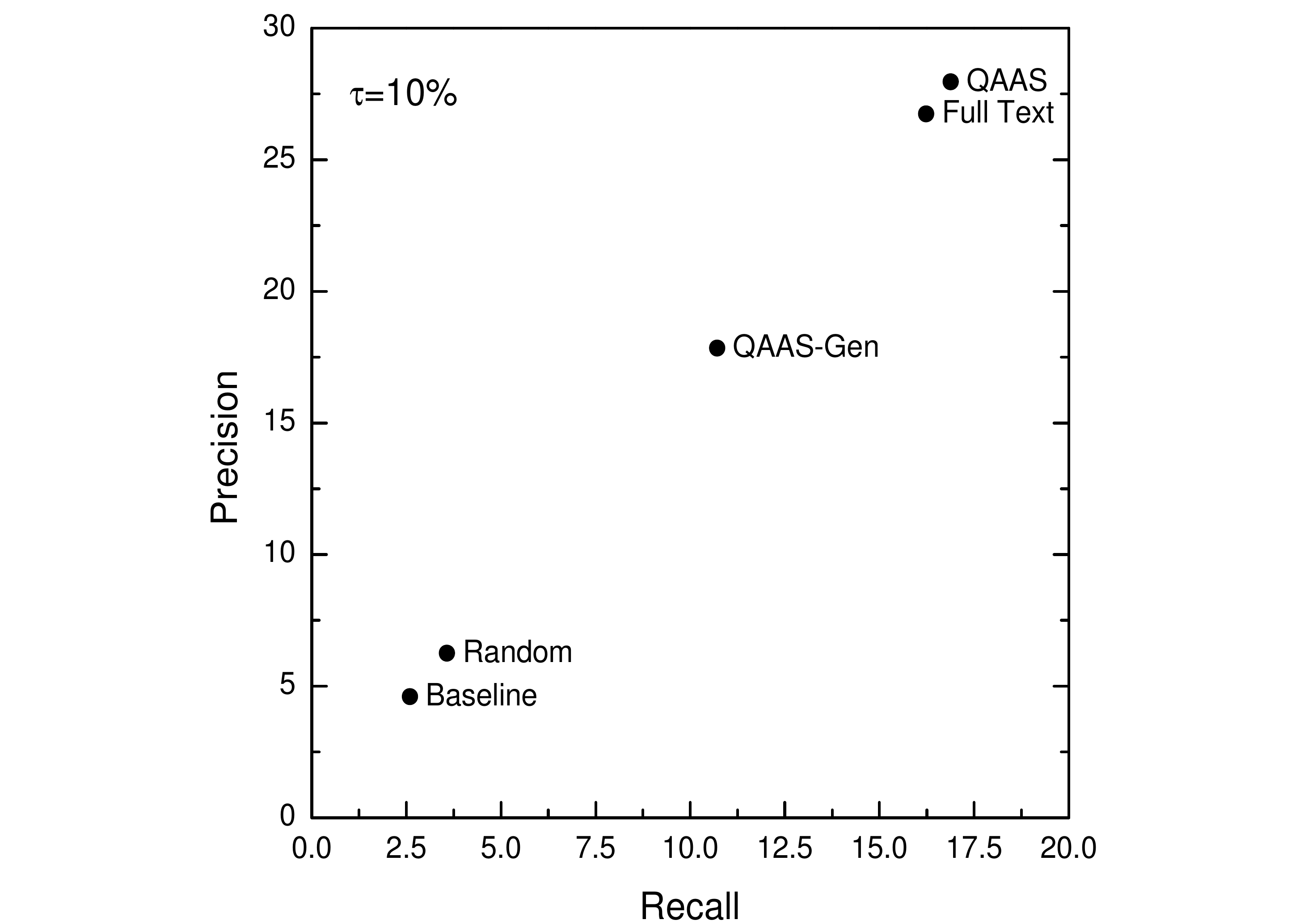}
             \includegraphics[width=18pc]{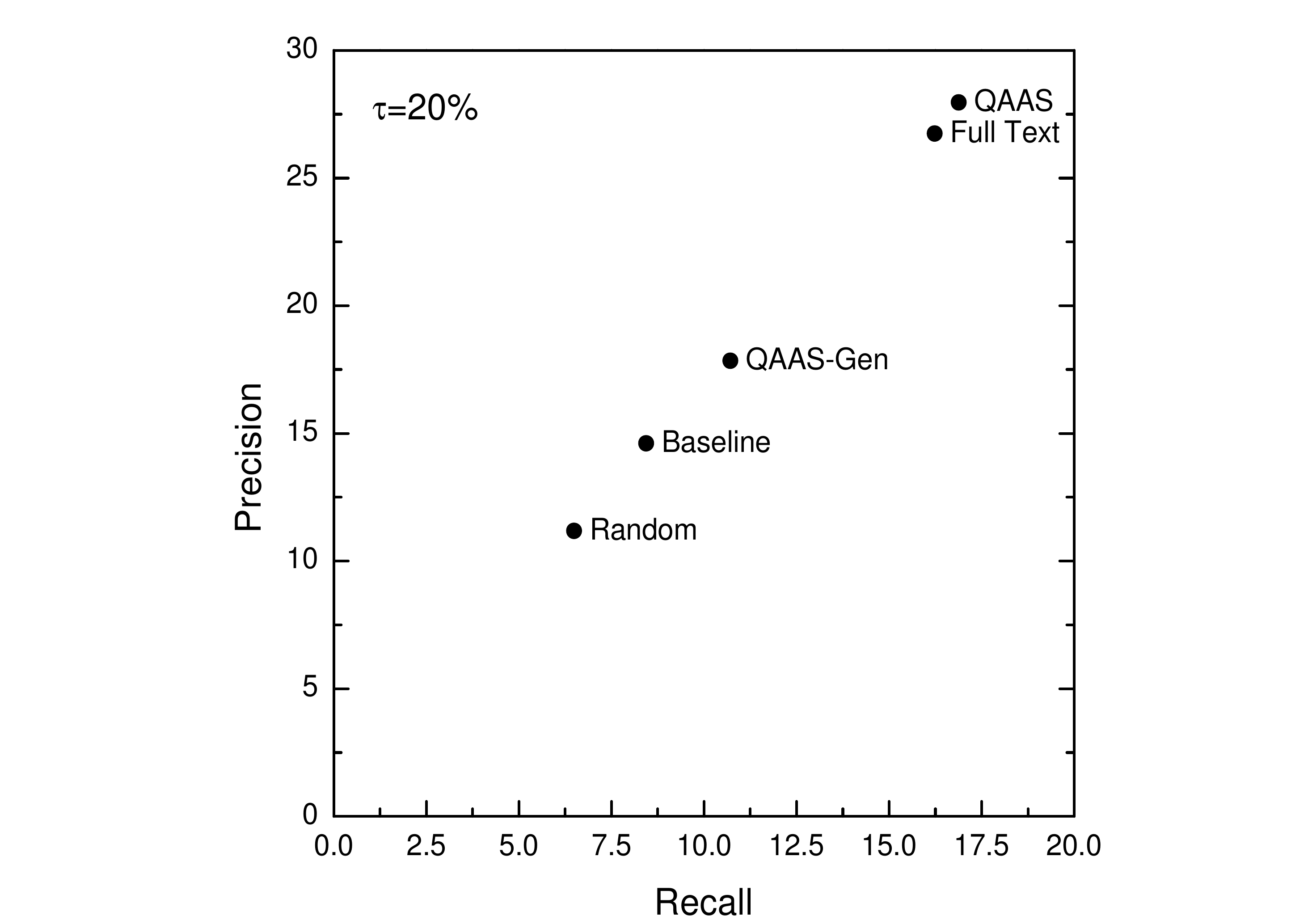}}
 \caption{\emph{Precision-Recall values for all  systems and for full text research (from table \ref{tab:EN-custom}). 
          For summarization systems, we fixed the compression rate $\tau = 10\%$ (left)
          and $\tau = 20\%$ (right).
          Systems on the top and right are better. The performance of the personalized QAAS 
          system is the best, and the volume of the search space is less important. 
          }}
  \label{fig:CWS-Qual}
\end{figure}

Figures \ref{fig:CWS-PERS-GEN} and \ref{fig:CWS-Qual}  show our results. 
We have compared the QAAS system with the generic {\sc Cortex} summarization system and the Baseline system. Baseline tests were
performed in two ways: as a random system or a baseline system (the first percent of sentences). In both cases, 
the score for each sentence (a value in [0,1]) was randomly generated.

Figure \ref{fig:CWS-PERS-GEN} shows a comparison between baseline/random methods and
personalized extracts from the {\sc Cortex} system. We note that
personalized summaries are much better than other methods.
However, the baseline system obtains a good performance at compression rates lower than 50\%. This can be explained
by the nature of the corpus. The newspaper articles are written in an "intrinsic" baseline form, 
the common style of journalists: the main information is duplicated and located at the top (first lines) of document.

Finally, Figure \ref{fig:CWS-Qual} shows the  
Precision-Recall values (Correct answers, Responses) for the random and baseline systems, the QAAS system with generic and customized summarization, and QA applied to the full text corpus: the precision value for the personalized QAAS system
is higher than the precision value obtained with full text. 

\subsection{Analysis of the results}



The results obtained show that degradation is minimal between 1\% and 3\% in spite of a high compression rate, when customized summary is used. 
When full documents (without being summarized) are processed by our information retrieval system, the answers (the references) are found in 209 cases out of 308\footnote{For 20 questions there is no response.}. 
But in 48 cases, the segments associated with these answers do not correspond to the question (even if they contain the words of the answers, they are used in a context different from the context of the question). Thus, only 158 of the 206 answers found in the documents may be considered as correct. 
Let us give in example the following question: "Who has created The New York Post?"\footnote{« \textit{Qui a créé le New-York Post ?} »}. 
The correct answer \textbf{Alexandre Hamilton} is returned by the system, based on the following segment:

\textsl{"Created by the conservator \textbf{Alexandre Hamilton} in 1801, it remained faithful to the ideas of its founder but it often changed hands, particularly lately: briefly property of Rupert Murdoch in the Eighties, the \textbf{Post} was, since 1988, that of Mr. Peter Kalikow, inheritor of a real-estate empire"}\footnote{\textit{Créé en 1801 par le conservateur \textbf{Alexandre Hamilton}, il est resté fidèle aux idées de son fondateur mais il a souvent changé de mains, particulièrement ces
derniers temps : brièvement propriété de Rupert Murdoch dans les années 80, le \textbf{Post} était, depuis 1988, celle de M. Peter
Kalikow, l'héritier d'un empire immobilier}.}.

Here the correct reference was indeed found, but the sentence used by the system to find it is considered to be insufficient at the time of the  evaluation. The segment does mention the creation of a newspaper called \textbf{Post} by Alexandre Hamilton, but there is no evidence  that this one is The \textbf{New York} Post.


Here is a similar example: the correct answer \textbf{Mitch} to the question "What hurricane devastated Central America in 1998?"\footnote{« \textit{Quel ouragan a dévasté l'Amé\-rique centrale en 1998 ?} »} is  found by the system, but the justification is insufficient: Hurricane \textbf{Mitch} devastated Central America short time after Johnny "had set fire at {\slshape Stade de France}"\footnote{\textit{L'ouragan \textbf{Mitch} a dévasté l'Amérique centrale peu de temps après que Johnny eut « mis le feu
au Stade de France »}}. Since no date is mentioned in the passage, it cannot be considered as supporting the answer.

\section{Conclusion}

The {\sc Cortex} algorithm is a very powerful text summarization system. 
We measured the quality of our summaries with intrinsic and extrinsic
methods. 
In intrinsec evaluation methods, our digests have a similar or higher quality than other methods. 
Our algorithm is able to process large corpus in three language (English, French and Spanish). 
Balanced summaries are obtained, and the majority of topics are taken into account. 
The Decision Algorithm, based on the weighting votes of metrics, is robust, convergent and independent of the order of segments.  
Two extrinsic methods were used to evaluate the quality of summaries: 
we coupled generic and query guided text summarization systems with a
question-answering system. Generic summaries act as a powerful
noise filter, but the quantity of answers found by the Question-Answering (QA) system
is a decreasing function of the compression rate. However, with a
customized summary, where texts are filtered and condensed in a
targeted way, the QA system performs much better. Customized summaries
reduce the risk of eliminating correct answers. Tests on the corpus
{\slshape Le Monde} showed that the {\sc Cortex} algorithm preserves the relevant
sentences, and that the QA system preserves its good performance,
evaluated by $CWS$ criterion. This is true even at high compression
rates (about 10\%), when customized summary is used. 
We think that the number of correct answers may be increased
if the system calculates the most appropriate Named Entity in the summarizing step before invoking the QA system.
Currently we are in the process of improving our system with that feature.

\section*{Acknowledgments}
{\small 
The authors give to thank Peter Peinl for the English, Laurent Gillard for the corpus, Grégoire Moreau for his reading and assistance with perl and regexp and Rafael Vazquez for some technical tests.
We also thank Sinequa for providing the labeled corpus containing Named Entities and the translation of the questions. 
P.-L. St-Onge, M. Gagnon and J.-M. Torres-Moreno thank Natural Sciences and Engineering Research Council of Canada (NSERC).
}

\appendix 

\section{The text \textsl{"Puces"}}
\label{ann:puces}

Note that the sentence containing the fragment \textsl{"...stationnée à Avenches, sont envahis {\bfseries parles} puces..."} contains the following mistake: \textsl{{\bfseries parles}} (to speak) must be written \textsl{{\bfseries par les}} (by the).
The segment \textsl{"...Des piqûres de puces ont été relevées sur plus d'{\bfseries untiers} des..."} contains the following mistake: {\bfseries untiers}, it must be written {\bfseries un tiers} (a third party). However, a small quantity of noise does not to much affect the {\sc Cortex} algorithm performance.

\vspace{.3cm}

{\footnotesize \sl
\noindent {\sc Informatique et puces.} 

\noindent Et si l'ordinateur pouvait fonctionner un jour, sans électricité
ou presque ? La démarche de chercheurs américains de l'université de Notre Dame, dans l'Indiana, montre que l'on peut manipuler des électrons pour construire des  circuits élémentaires avec des quantités d'énergie infimes. Leurs expériences, relatées dans l'édition du 9 avril du magazine Science, ouvrent la voie à des composants capables de fonctionner à des fréquences 10 à 100 fois plus élevées que celles des puces actuelles qui sont bridées par des problèmes de dissipation de chaleur. Les travaux de l'équipe dirigée par Greg Snider portent sur le puits quantique, un piège infinitésimal dans lequel un électron peut être enfermé. Les scientifiques ont créé des cellules carrées formées de quatre puits quantiques, dans laquelle ils ont introduit une paire d'électrons. Les forces de répulsion provoquent le déplacement des électrons qui trouvent leur équilibre lorsqu'ils se trouvent placés aux deux extrémités de l'une ou l'autre des diagonales de la cellule. La première représente l'état 0, tandis que l'autre indique le 1: chaque cellule représente donc un bit, la plus petite quantité d'information que l'on peut manipuler dans les ordinateurs. Tout déplacement d'un électron sous l'effet d'une force extérieure provoque automatiquement le déplacement du second électron de manière à retrouver l'équilibre, et donc le basculement de la cellule entre les états 0 et 1. L'utilisation d'une cellule unique ne prouve rien. Les chercheurs américains ont réussi à en assembler plusieurs, provoquant, suivant leurs besoins, le déplacement des électrons sans devoir fournir d'énergie, ou presque. Dans les transistors actuels, le passage de l'état 0 à l'état 1 n'est possible qu'au prix du déplacement de plusieurs milliers d'électrons, ce qui génère un important flux de chaleur. En regroupant cinq cellules élémentaires, les chercheurs ont mis au point un circuit baptisé "majoritaire" capable de réaliser les deux fonctions logiques de base, ET et OU, à la demande. Ils ont ensuite vérifié son bon fonctionnement et espèrent assembler plusieurs de ces circuits pour effectuer des additions et des multiplications sur des nombres. En cas de succès, la technique des cellules logiques quantiques pourrait permettre d'entasser des centaines de milliards de circuits dans une seule puce électronique. Pour l'instant, le dispositif fonctionne seulement à une température voisine du zéro absolu, mais les chercheurs ne désespèrent pas de parvenir à le réchauffer tout en maîtrisant son comportement.
Les cantonnements de la compagnie IV de l'école de recrues
d'infanterie d'explora\-tion et de transmission 213, stationnée à
Avenches, sont envahis par les puces et les poux. Des piqûres de
puces ont été relevées sur plus d'untiers des militaires. On a
aussi retrouvé des cadavres de poux sur 3 militaires. Des mesures
d'urgence ont été prises en conséquence. Des piqûres de puces ont
été diagnostiquées sur plus d'un tiers des 155 hommes de la
compagnie IV de l'école de recrues d'infanterie d'exploration et
de transmission 213. Des cadavres de poux, mais aucun oeuf, ont
également été décelés sur 3 militaires. Ces insectes sont transmis
par contact personnel. La cause de cette invasion n'est pas
claire; ces insectes semblent toutefois avoir essaimé à partir du
local de garde. Le médecin de troupe a donné immédiatement les soins
nécessaires aux militaires concernés et il a ordonné les mesures d'hygiène
qui s'imposaient. Des produits spéciaux ont été remis pour les soins corporels.
Tout le matériel personnel delà compagnie a été emballé hermétiquement et
apporté à l'arsenal cantonal de Fribourg. La troupe sera déplacée dans un
complexe industriel. Une section d'hygiène de l'école de recrues d'hôpital 268,
stationnée à Moudon, va désinfecter tous ces cantonnements. On estime qu'avec
ces mesures sanitaires appropriées la troupe pourra réintégrer ses cantonnements vendredi au plus tard.
}

\subsection*{Generic abstract generated by {\sc Cortex} ($\tau=25\%$, in brackets the number of extracted sentences)}

\noindent {\small $^{[1]}$La démarche de chercheurs américains de l'université de Notre Dame, dans l'India\-na, montre que l'on
peut manipuler des électrons pour construire des circuits élémen\-taires avec des quantités d'énergie infimes. {\small $^{[5]}$Les forces de répulsion provoquent le déplacement des électrons qui trouvent leur équilibre lorsqu'ils se trouvent
placés aux deux extrémités de l'une ou l'autre des diagonales de la cellule. {\small $^{[8]}$Tout déplace\-ment d'un électron
sous l'effet d'une force extérieure provoque automatiquement le déplacement du second électron de manière à retrouver
l'équilibre, et donc le basculement de la cellule entre les états 0 et 1. {\small $^{[10]}$Les chercheurs américains ont réussi
à en assembler plusieurs, provoquant, suivant leurs besoins, le déplacement des électrons sans devoir fournir d'énergie,
ou presque. {\small $^{[12]}$En regroupant cinq cellules élémentai\-res, les chercheurs ont mis au point un circuit baptisé
"majoritaire" capable de réa\-li\-ser les deux fonctions logiques de base, ET et OU, à la demande. {\small $^{[14]}$En cas de succès, la technique des cellules logiques quantiques pourrait permettre d'entasser des centaines de milliards de circuits
dans une seule puce électronique. {\small $^{[16]}$Les cantonne\-ments de la compagnie IV de l'école de recrues d'infanterie
d'exploration et de transmission 213, stationnée à Avenches, sont envahis parles puces et les poux. {\small $^{[20]}$Des piqûres
de puces ont été diagnostiquées sur plus d'un tiers des 155 hommes de la compagnie IV de l'école de recrues d'infanterie
d'exploration et de transmission 213.
}

\bibliographystyle{splncs}
\bibliography{qaas}

\end{document}